\documentclass[a4paper,11pt]{article}

\usepackage{jheppub}

\usepackage[T1]{fontenc}

\usepackage{wasysym}

\usepackage{comment}
\usepackage{graphicx}
\usepackage{amsmath}
\usepackage{amsfonts}
\usepackage{amssymb}
\usepackage{mathrsfs}
\usepackage{natbib}
\usepackage{xcolor}
\usepackage[colorlinks=true
,urlcolor=blue
,anchorcolor=blue
,citecolor=blue
,filecolor=blue
,linkcolor=blue
,menucolor=blue
,linktocpage=true
,pdfproducer=medialab
,pdfa=true
]{hyperref}
\usepackage{color}
\usepackage{comment}
\usepackage{xcolor}
\usepackage{cleveref}
\usepackage[utf8]{inputenc}
\usepackage{subcaption}
\usepackage{booktabs}

\newcommand{\beq}{\begin{equation}}
\newcommand{\eeq}{\end{equation}}
\newcommand{\bea}{\begin{eqnarray}}
\newcommand{\eea}{\end{eqnarray}}
\newcommand{\bef}{\begin{figure}}
\newcommand{\eef}{\end{figure}}

\title{\Large Exotic Higgs Decays at a Muon Collider}

\author[a,b]{JiJi Fan,}
\author[c,d]{Lingfeng Li,}
\author[e,f]
{Tao Liu,}
\author[a]
{Yanhan Wang,}
\author[e]{and Mingrui Zhou}
\affiliation[a]{Department of Physics, Brown University, Providence, RI 02912, USA}
\affiliation[b]{Brown Center for Theoretical Physics and Science and Innovation, Brown University, Providence, RI 02912, USA}
\affiliation[c]{International Center of Theoretical Physics-Asia Pacific,\\ University of Chinese Academy of Sciences, Beijing 100190, China}
\affiliation[d]{Institute of High Energy Physics, Beijing 100049, China}
\affiliation[e]{Department of Physics, 
The Hong Kong University of Science and Technology, Hong Kong S.A.R., China}
\affiliation[f]{Jockey Club Institute for Advanced Study,
The Hong Kong University of Science and Technology, Hong Kong S.A.R., China}

\abstract{We study the sensitivity of a future muon collider to exotic Higgs decays in a minimal scenario of Standard Model (SM) augmented with a light singlet scalar $S$. We consider the decay $h\to SS$ and $S$'s subsequently decay back to SM. In particular, we focus on final states with four bottom quarks ($4b$), and two bottom quarks and two muons ($2b2\mu$). Analyses are performed for two muon collider benchmark configurations: center-of-mass collision energy $\sqrt{s}=3~\mathrm{TeV}$ with $1~\mathrm{ab}^{-1}$ data and $\sqrt{s}=10~\mathrm{TeV}$ with $10~\mathrm{ab}^{-1}$ data. Machine-learning techniques are applied to suppress backgrounds and mitigate jet-combinatorics effects in both channels. We find that the $4b$ mode could be sensitive to the branching ratio, BR$(h \to SS \to 4b)$, of ${\cal O}(10^{-2})$ at 3~TeV and ${\cal O}(10^{-3})$ at 10~TeV, significantly improving upon high-luminosity LHC projections. In the Higgs-portal model with $S$ coupling to SM only through mixing with the Higgs, the sensitivities to BR$(h \to SS)$ remain at the same level given ${\cal O}(1)$ branching fraction of $S$ decaying into $b$-quarks.
The $2b2\mu$ mode benefits from a clean dimuon resonance and can probe BR$(h\to SS\to 2b2\mu)$ down to $10^{-5}$ level at a $10~\mathrm{TeV}$ muon collider. But the sensitivity to BR$(h \to SS)$ will be significantly reduced due to the small branching fraction of $S$ decaying into muons in the Higgs portal model. }

\begin{document}

\maketitle

\section{Introduction}
\label{sec:intro}

The discovery of the Higgs boson at the Large Hadron Collider (LHC) marks the completion of the Standard Model (SM) and starts a new chapter for particle physics. Since then, the Higgs boson has become a key experimental target: the precision measurements of its properties are among the top priorities at the collider frontier in the foreseeable future. One outstanding opportunity in the Higgs program is the search for exotic Higgs decays, in which Higgs decays to new light particles beyond the SM. Such exotic decays appear in a large variety of new physics scenarios, driven by some deepest questions in particle physics including naturalness, dark matter, and electroweak phase transitions (EWPT).  It has long been known that exotic Higgs decays serve as powerful probes to new physics~\cite{Shrock:1982kd} and their theoretical studies have picked up a higher momentum after the Higgs discovery (see~\cite{Curtin:2013fra} for a comprehensive study, and~\cite{Cepeda:2021rql} for a review). On the experimental side, the large samples of the Higgs bosons that have been and will be produced at the LHC allow us to test different theoretical possibilities of exotic decays directly, in particular in the upcoming high-luminosity runs.

Beyond the LHC, the community has been actively discussing possible future colliders to take over the barton of new physics searches. Among different choices, a future high-energy muon collider offers a unique combination of being a high-precision and a high-energy machine simultaneously. As the colliding muons are elementary particles, a muon collider provides a cleaner environment compared to more noisy machines colliding composite hadrons, and enables precision measurements. On the other hand, since muons are much heavier than electrons, synchrotron radiation in circular motions of muons is much more suppressed than that of electrons, allowing a circular muon collider to achieve a much higher center-of-mass collisional energy and become a direct discovery machine. Due to these advantages, there has been a growing interest in investigating the potential of a muon collider in different aspects, such as measuring the SM Higgs properties~\cite{Forslund:2022xjq,deBlas:2022aow,Forslund:2023reu,Li:2024joa,Chen:2021pqi,Chen:2022yiu,Celada:2023oji} or other SM processes~\cite{Azatov:2022itm,Yang:2020rjt,Yang:2022fhw,Fridell:2023gjx,Ma:2024ayr,Dong:2023nir,Altmannshofer:2022xri,Zhang:2023yfg,Han:2023njx,Zhang:2023khv,Han:2024gan}, and searching for various new physics scenarios~\cite{Liu:2021akf,Li:2023tbx,Kwok:2023dck,Chen:2022msz,Cesarotti:2022ttv,Bao:2022onq,Li:2021lnz,Dermisek:2021mhi,Homiller:2022iax,Sen:2021fha,Dasgupta:2023zrh,Jueid:2023zxx,Haghighat:2021djz,Casarsa:2021rud,Cesarotti:2023sje,Jueid:2023qcf,Das:2023tna,Li:2023lkl,Black:2022qlg,Ghosh:2023xbj,  Bandyopadhyay:2024plc,Bandyopadhyay:2024gyg, Lu:2023ryd,Mikulenko:2023ezx,Liu:2023jta,Li:2022kkc,Chigusa:2023rrz,Medina:2021ram,Han:2025wdy,Han:2022edd,Han:2022ubw,Han:2022mzp,Han:2021udl,Jana:2023ogd,Barducci:2024kig,He:2024dwh,Cao:2024rzb,Bi:2024pkk,Dehghani:2025xkd,Ghosh:2025dcv,Saha:2025npi,Chakraborty:2022pcc,Coleppa:2026esg}. For reviews and community reports, see~\cite{AlAli:2021let,Accettura:2023ked,Aime:2022flm,Black:2022cth,InternationalMuonCollider:2024jyv,Acosta:2022ejc}. 

 One aspect which has not been fully explored is the prospect of probing exotic Higgs decays at a future muon collider. This will be the focus of our paper. More specifically, we focus on one classic benchmark scenario in which the Higgs decays to a pair of SM-gauge-singlet scalars, which subsequently decay back to the SM~\cite{OConnell:2006rsp, Profumo:2007wc,Curtin:2014pda, Kozaczuk:2019pet,Carena:2019une,Shelton:2021xwo,Adhikary:2022jfp,Liu:2022nvk,Wang:2022dkz,Carena:2022yvx,Roche:2023cun,Roche:2023int,Yu:2024xsy,Cheng:2024gfs,Hammad:2024hhm,Li:2025luf,DAgnolo:2025cxb}. This could lead to fully hadronic, semi-leptonic, and full leptonic final states. Though a muon collider does not show an advantage over the high-luminosity LHC (HL-LHC) in the full leptonic channel, we will show that it could improve the sensitivity significantly in the full hadronic channel such as four bottom-quark final state, and semi-hadronic channel, e.g., final state of two bottom quarks plus two muons. 

 The paper is organized as follows. In Sec.~\ref{sec:models and simulations}, we will review the model in which the Higgs boson could decay to two singlet scalars beyond the SM and describe the simulation procedures for both the singals and associated SM backgrounds. In Sec.~\ref{sec:analysis}, we will present details of the analysis in which we apply machine learning techniques and discuss the key results. We will conclude and outline future directions in Sec.~\ref{sec:con}.

\section{Models and Simulations}
\label{sec:models and simulations}
In this section, we first review a benchmark model which allows the Higgs boson to decay to a pair of singlets beyond the SM. This will be the main exotic Higgs decay scenario we focus on. Then we will describe in detail the simulation setup for both the new physics signals and their relevant SM backgrounds. 

\subsection{Model}

We consider a minimal extension of the SM in which the Higgs boson couples to and mixes with a SM-gauge-singlet real scalar field, $S$. The Higgs-scalar interaction potential is given by~\cite{OConnell:2006rsp,Profumo:2007wc,Kozaczuk_2020,Wang:2023zys}:
\begin{equation}
V = -\mu^2|H|^2+ \lambda |H|^4+\frac{1}{2}a_1|H|^2S + \frac{1}{2} a_2|H|^2 S^2 + b_1 S+ \frac{1}{2}b_2 S^2 + \frac{1}{3}b_3S^3+\frac{1}{4}b_4S^4 \, ,
\label{interaction-potential}
\end{equation}
where $H$ refers to the SM Higgs doublet field. $\mu^2$ and $\lambda$ correspond to the Higgs mass squared parameter and quadratic coupling, respectively. The coefficients $a_1$ and $a_2$ describe the interaction between the Higgs doublet and the scalar singlet, with $a_1$ inducing the Higgs–singlet mass mixing after electroweak symmetry breaking (EWSB). The remaining parameters $ b_1$, $ b_2$, $ b_3$, and $b_4$ govern the singlet-sector potential, controlling the singlet vacuum expectation value (VEV), mass, and self-interaction. Together, these terms constitute the most general renormalizable scalar potential involving $S$ and $H$. After EWSB, the two fields could be parametrized as 
\begin{equation}
H = \frac{1}{\sqrt{2}}\begin{pmatrix} 0 \\ v + h \end{pmatrix}, 
\qquad
S = v_s + s \, ,
\end{equation}
where $v = 246 $ GeV is the VEV of the Higgs field $H$ and $v_s$ is the VEV for $S$. The gauge-singlet scalar may be shifted by a constant without altering physical observables, as it couples to other SM fields only through the Higgs field. We therefore work in the $v_s = 0$ basis.
The two scalar fields $h$ and $s$ mix, and the corresponding mass eigenstates are given by:
\begin{equation}
\begin{aligned}
h_1 &= h \cos\theta + s \sin\theta ,\\
h_2 &= -\,h \sin\theta + s \cos\theta \, ,
\label{define_h1h2}
\end{aligned}
\end{equation}
where $h_1$ denotes the singlet-like mass eigenstate with a mass $m_1$, while $h_2$ corresponds to the Higgs particle with $m_2 \approx 125$ GeV. $\theta$ is the mixing angle.  The trilinear scalar interaction can be written in terms of mass eigenstates as:
\begin{equation}
V \supset 
\frac{1}{6}\,\lambda_{111}\,h_1^3
+ \frac{1}{2}\,\lambda_{211}\,h_2 h_1^2
+ \frac{1}{2}\,\lambda_{221}\,h_2^2 h_1
+ \frac{1}{6}\,\lambda_{222}\,h_2^3 \, .
\end{equation}
The coefficients $\lambda_{ijk}$'s denote the trilinear couplings among the $i$th, $j$th and $k$th scalar mass eigenstates. Specifically, $\lambda_{111}$ and $\lambda_{222}$ correspond to the self-interactions of the singlet-like scalar $h_1$ and the Higgs boson $h_2$ respectively, while $\lambda_{221}$ describes interaction involving two Higgs bosons and one singlet-like scalar. The coupling $\lambda_{211}$ is of special phenomenological importance, as it governs the interaction between one Higgs boson and two singlet-like scalars and directly controls the exotic Higgs decay process $h_2 \to h_1 h_1$. 
The partial width of this decay is given by
\begin{equation}
\Gamma\!\left(h_2 \to h_1 h_1\right)
=
\frac{1}{32\pi\,m_2}\,
\lambda_{211}^2\,
\sqrt{1 - \frac{4 m_1^2}{m_2^2}} \, .
\label{eq:partialwidth}
\end{equation}
The decay is kinematically allowed only if $m_1 < m_2/2$.

In the small-mixing limit, $\theta \ll 1$, where the mass eigenstates $h_1$ and $h_2$ defined in Eq.~\eqref{define_h1h2} are dominantly the singlet scalar and the SM-like Higgs boson, respectively. For phenomenological convenience and given that the existing data is consistent with the Higgs boson being SM-like, we work in this limit and therefore identify $h_2 \equiv h$ with mass $m_h \approx 125$ GeV and $h_1 \equiv S$ ($S=s$ in the $v_s=0$ basis) with mass $m_S$ throughout the remainder of this work. In this case, Eq.~\eqref{eq:partialwidth} describes the exotic Higgs decay $h \to SS$.
The singlet scalar $S$ inherits Higgs-like couplings to other SM particles through $S$-$h$ mixing. Thus the decay modes of $S$ have their partial widths as those of the Higgs boson at the same mass times $\theta^2$~\cite{Fradette:2017sdd}. In particular, for the mass range of $S$ we are interested in between 10 and 60 GeV, $S$ decays mostly to SM fermions and the dominant channel is $S \to b\bar{b}$ with $b$ bottom quarks. We include leading QCD corrections in the computations of the corresponding partial hadronic decay widths~\cite{Drees:1990dq}.   


%

\subsection{Simulation Setups}
We use \textsc{Madgraph~5}~\cite{Alwall:2014hca} to generate parton-level processes for both exotic Higgs decay signals and associated SM backgrounds. Parton and electromagnetic showering are simulated using \textsc{Pythia~8}~\cite{Sjostrand:2014zea}. We use \textsc{Delphes~3}~\cite{deFavereau:2013fsa} for detector simulation of a muon collider. We consider two muon collider benchmarks with center-of-mass energy at 3~TeV and 10~TeV, and integrated luminosity of 1~ab$^{-1}$ and 10~ab$^{-1}$ respectively.

When simulating new physics signals, we implement the model in \textsc{Feynrules}~\cite{Alloul:2013bka} and then import it to \textsc{Madgraph~5}. We focus on two final states 4$b$ (4 bottom quarks) and 2$b$2$\mu$ (2 bottom quarks plus 2 muons), which are representative decay modes of the singlet scalar. For the benchmark model with $m_S$ in the range of (10 - 60) GeV, $h \to 2S \to 4b$ is the dominant exotic decay channel. We also explore the semi-leptonic channel $h \to 2S \to 2b 2\mu$, considering its relative cleaner background. The full leptonic decay modes $h \to 2 S \to 4e, 4\mu, 2e2\mu$ are significantly suppressed in the benchmark model. In addition, we find through simulations and detailed analyses that the sensitivity of a muon collider to the full leptonic final state does not improve over that of the near-future HL-LHC in general. For both decay channels, we simulate signals with six different $m_S$ benchmarks: $m_{S} = 15, 20, 30, 40, 50, 60$ GeV.

The dominant Higgs production channel for both the signal and the relevant SM background is vector boson fusion (VBF). In particular, the charged-current process \(\mu^+ \mu^- \to \nu \bar{\nu}h\) mediated by $W$-boson fusion gives the leading contribution with a cross section of approximately $1~\mathrm{pb}$ at $\sqrt{s}=10~\mathrm{TeV}$.
The neutral-current channel \(\mu^+ \mu^- \to \mu^+ \mu^- h\) arising from $Z$/$\gamma$ fusion is subdominant, with a cross section of $\sim 0.1~\mathrm{pb}$ when $\sqrt{s}=$10 TeV. We therefore focus on $W$-boson fusion in our analysis. 

At parton-level, all $b$ quarks are required to have transverse momenta $p_T^b > 15~\mathrm{GeV}$, and pseudorapidity $|\eta^b| < 2.5$ for both signal and background generation, consistent with the muon-collider detector acceptance and to ensure stable and efficient event generation. To eliminate configurations in which two partons are collinear and would be misidentified into a single jet, an angular separation requirement on any two $b$ quarks, $\Delta R_{bb} > 0.25$, is imposed. For processes with leptons ($i.e.$ muons) in the final state, such as $2b2\mu$, basic lepton (denoted by $\ell$) acceptance cuts are applied at the generator level, including requirements on the lepton's transverse momentum $p_T^\ell >0.5~\mathrm{GeV}$ and its pseudorapidity $|\eta^{\ell}| < 8.0$.
Requirements of minimum separation between two leptons as well as between a $b$ quark and a lepton, $\Delta R_{\ell\ell} > 0.1$ and $\Delta R_{\ell b} > 0.25$, are applied to suppress collinear configurations and overlapping muon–jet topologies.

Detector effects are simulated using the default muon collider detector template~\cite{deFavereau:2013fsa}. Jets are reconstructed with the Valencia (VLC) algorithm~\cite{Boronat:2014hva} and a radius parameter = 0.5. We adopt the $b$-tagging working point with a flat $b$-tagging efficiency of 70\%. To improve the reliability of jet-flavour association in dense hadronic environments, the cone of $b$-flavor matching between reconstructed jets and the $b$ quark is reduced to $\Delta R < 0.3$. This choice leads to a modest reduction in the overall $b$-tagging performance in multi-jet final states but significantly suppresses mis-tagging caused by overlapping or adjacent $b$-hadrons. In particular, it mitigates configurations in which multiple reconstructed jets are accidentally associated with the same $b$-hadron. We also apply a correction to the energies of $b$-tagged jets in all samples, following a rough approximation adopted by an ATLAS study~\cite{ATLAS:2017cen}. This correction is intended to mitigate various energy losses such as the one due to invisible neutrinos from semi-leptonic $b$-hadron decays. For channels containing muons, only muons with $|\eta^\mu| < 2.5$ will be selected for event resconstruction.


\subsubsection{Final State : $4b$ }
We first consider the fully hadronic final state with the signal mode:
\begin{equation}
 h \to S S\to 4b \, .
\end{equation}
The relevant SM background can be categorized as Higgs-induced and non-Higgs processes. The dominant Higgs-induced background is $h \to ZZ^* \to 4b$ with $h$ mostly produced from the $W$-boson fusion process, which leads to an analogous and irreducible final state as the signal. 
Processes $h \to 4b$ without an intermediate on-shell $Z$ boson yield the same final state, but are subdominant because they lack the resonant enhancement from the on-shell $Z \to b\bar{b}$ decay. Other Higgs-induced background including a final state of $b \bar{b} j j$ with $j$ light jets could be significantly reduced and become numerically negligible after applying a 4$b$-tagging cut. All Higgs-induced background samples are normalized using the Higgs production cross sections and corresponding decay branching ratios in the SM~\cite{LHCHiggsCrossSectionWorkingGroup:2013rie} as the new physics corrections are highly suppressed.

The non-Higgs backgrounds include $\mu^+ \mu^- \to \nu_\mu \bar{\nu}_\mu\, b\overline{b}$ and $\mu^+ \mu^- \to \nu_\mu \bar{\nu}_\mu (Z^{(*)} Z^{(*)} \to 4b)$. For the first one $\mu^+\mu^- \to \nu_\mu \bar{\nu}_\mu b \overline{b}$, an invariant-mass cut of $100~\mathrm{GeV} < m_{bb} < 200~ \mathrm{GeV}$ is applied at the generator level to suppress the on-shell $Z\to b\bar b$ contribution and to select kinematic configurations similar to Higgs-induced $4b$ final state. This process does not naturally produce four hard 
$b$-jets, and additional $b$-jets from parton showers are typically soft or collinear, causing most background events from this process to fail the resolved four-$b$ selection and invariant mass requirements. The $\mu^+\mu^- \to \nu_\mu \bar{\nu}_\mu(Z^{(*)}Z^{(*)} \to 4b)$ background is also suppressed due to its smaller production cross section and the absence of a Higgs mass resonance in the 4$b$ system. 

\subsubsection{Final State : $2b2\mu$ }
We also consider the semi-leptonic final state. For the signal, we have
\begin{equation}
 h \to S S\to 2b2\mu \, .
\end{equation}
The background simulation is considerably simpler compared to 4$b$ final state, with Higgs-induced background: $\mu^+ \mu^- \to \nu_\mu \bar{\nu}_\mu (h \to Z Z^* \to 2b2\mu)$ and non-Higgs background: $\mu^+ \mu^- \to  \nu_\mu \bar{\nu}_\mu 2b2\mu$. The Higgs-induced background is further suppressed after applying invariant mass cuts on both $b\bar{b}$ and $\mu^+\mu^-$ pairs, as we will show in Sec.~\ref{sec:analysis}. For the non-Higgs background, we require the invariant mass of the muon pair to satisfy $10~\mathrm{GeV} < m_{\mu\mu} < 70 ~\mathrm{GeV}$  at the generator level, which greatly suppresses contributions, keeping only the kinematic region relevant for our $m_S$ benchmarks. We also simulate processes with a $\mu^+ \mu^- j j$ final state, where $j$ represents light jets. Owing to the small light-jet mis-tag probability, this channel yields a negligible contribution after $b$-tagging selection.

\section{Analysis and Results}
\label{sec:analysis}

In this section, we will present details of analyses after simulation and results for the two final states discussed in the previous section. 
In all the figures and tables throughout this section, we will write $h$ explicitly in labels of the Higgs-induced backgrounds while only indicate final state particles for the non-Higgs backgrounds. We will also not explicitly indicate neutrinos when labeling different backgrounds. 

\subsection{Preselection}
\label{sec:preselection}

\begin{figure}[h]
    \centering
    \begin{subfigure}[h]{0.48\textwidth}
        \centering
        \includegraphics[width=\textwidth]{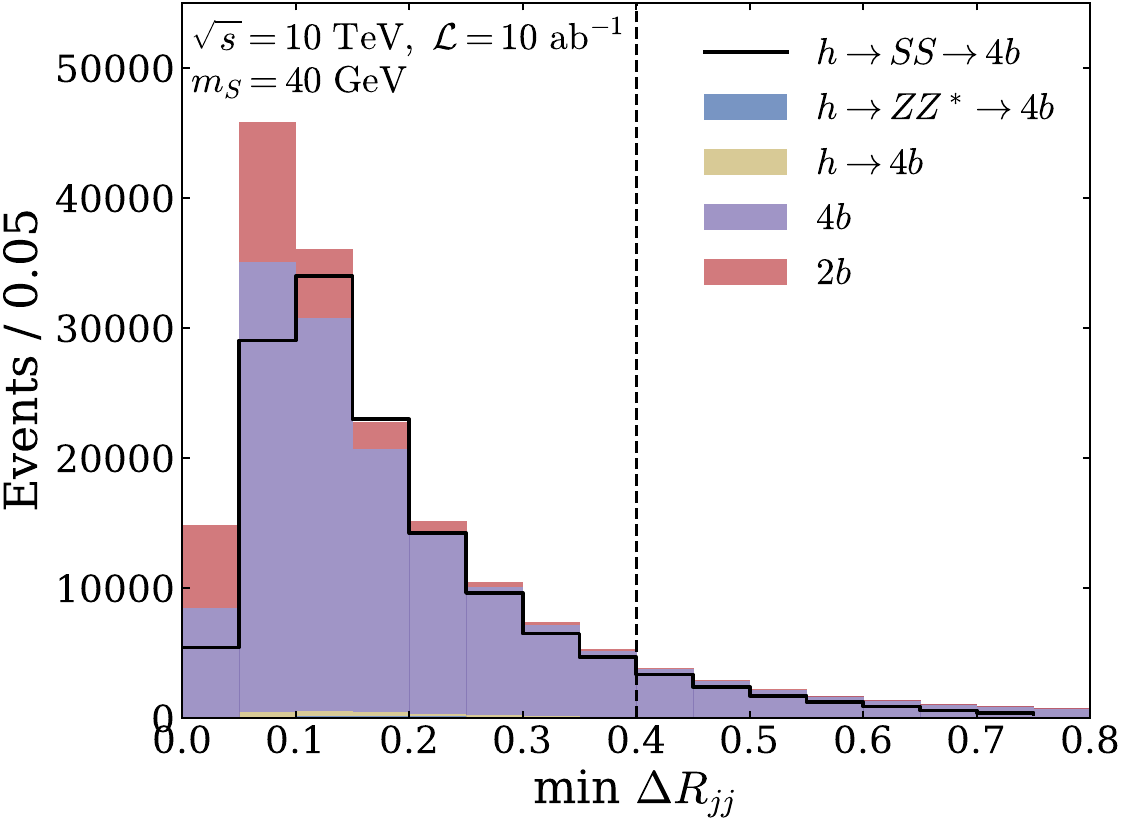}
        \label{fig:4b_dr}
    \end{subfigure}
    \hfill
    \begin{subfigure}[h]{0.48\textwidth}
        \centering
        \includegraphics[width=\textwidth]{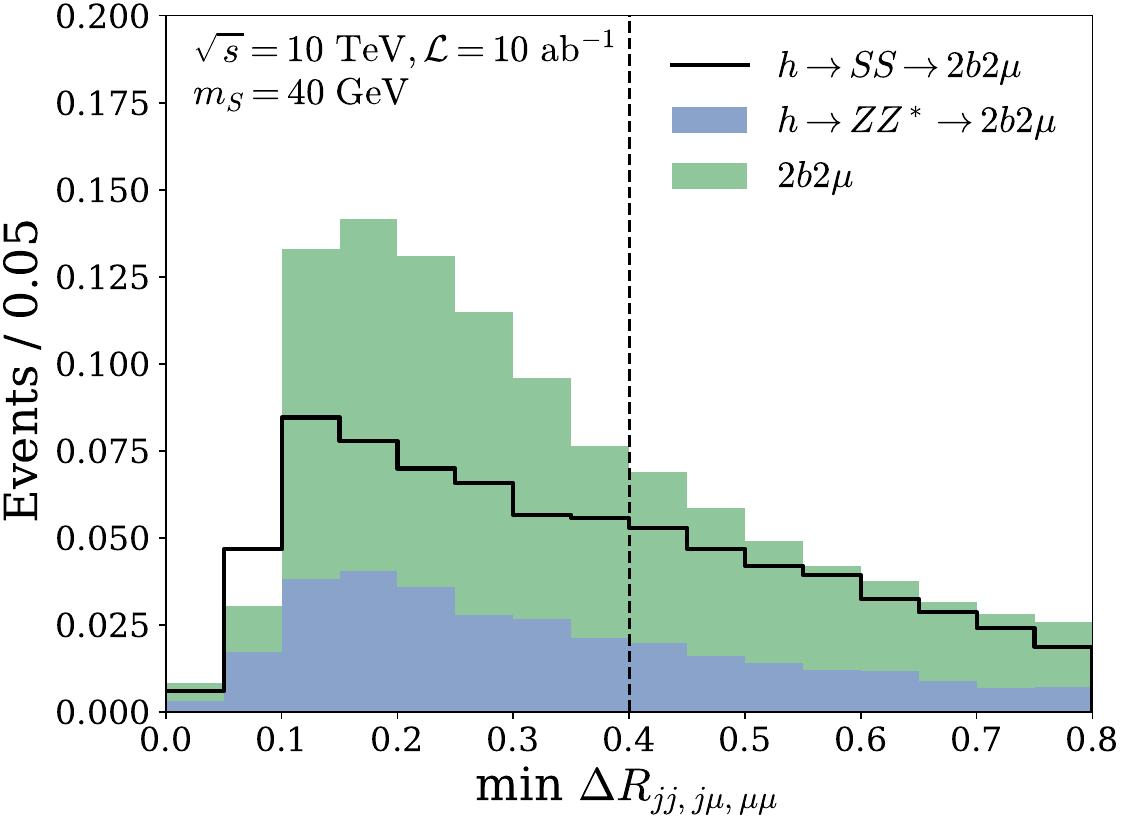}
        \label{fig:2b2m-dr}
    \end{subfigure}
    \caption{Distributions of the minimum pairwise angular separation, $\min  \Delta R$, for the $4b$ (left) and $2b2\mu$ (right) final states before imposing the $\Delta R$ requirements. For $4b$, $\min \Delta R$ is defined as the minimum angular separation among all jet pairs, while for $2b2\mu$, it is defined as the minimum separation among all $jj$, $\mu\mu$, and $j\mu$ pairs. The black solid lines indicate the signal distributions while histograms of different colors represent various leading SM backgrounds. The dashed vertical lines indicate the cuts of $\Delta R > 0.4$. Both panels show the scenario of a muon collider operating at $\sqrt{s}=10~\mathrm{TeV}$ with an integrated luminosity of $10~\mathrm{ab}^{-1}$. For the signal, we assume $m_S=$ 40 GeV. The signal yields are calibrated to branching ratios of $\mathrm{BR}(h\to SS\to 4b)=0.1$ and $\mathrm{BR}(h\to SS\to 2b2\mu)=5\times 10^{-7}$, which make the signal and background yields comparable.  }
    \label{fig:4b_dR_comparison}
\end{figure}

To suppress the SM backgrounds while maintaining a high signal efficiency, a series of preselection cuts is applied first. 
Similar preselection cuts are also applied to the $2b2\mu$ channel. We first require each reconstructed jet to have its transverse momentum $p_{T}^j > 20 ~\mathrm{GeV}$, which suppresses soft QCD radiation and low-energy jets that are poorly reconstructed. Note that this cut, as well as other preselection cuts on jets, are applied on all jet flavors disregard of $b$-tagging results. Compared to the parton-level cuts on transverse momentum, this requirement further ensures that jets lie within the efficient operating region of the detector and the $b$-tagging algorithm. In the $2b2\mu$ channel analysis, an additional requirement is imposed on muons, requiring their transverse momenta to satisfy $p_{T}^{\mu} >5~\mathrm{GeV}$. Based on the parton-level requirement $\Delta R_{bb} > 0.25$ and jet-flavour association radius of 0.3, we further impose an angular separation cut $\Delta R_{jj} > 0.4$. 
For the $2b2\mu$ final state, we further require $\Delta R_{\mu\mu}$ and $\Delta R_{j\mu}$ to be above 0.4. Distributions of the minimum pairwise angular separation for both final states studied before applying the $\Delta R$ cuts is presented in Fig.~\ref{fig:4b_dR_comparison}.  
Such $\Delta R$ cuts are necessary for two reasons. Firstly, they removes QCD radiated collinear jets and thus strongly suppress two-$b$ background, as shown in the left panel of Fig.~\ref{fig:4b_dR_comparison}. Moreover, the moderate min$\Delta R_{jj}$ requirement significantly mitigates the mistagging of $b$-jets from the \texttt{Delphes} jet flavor association process. Numerically, we find that events without four parton-level $b$ quarks become negligible after the $\Delta R$ cuts are applied.

\begin{figure}[htbp]
    \centering
    \begin{subfigure}[h]{0.48\textwidth}
        \centering
        \includegraphics[width=\textwidth]{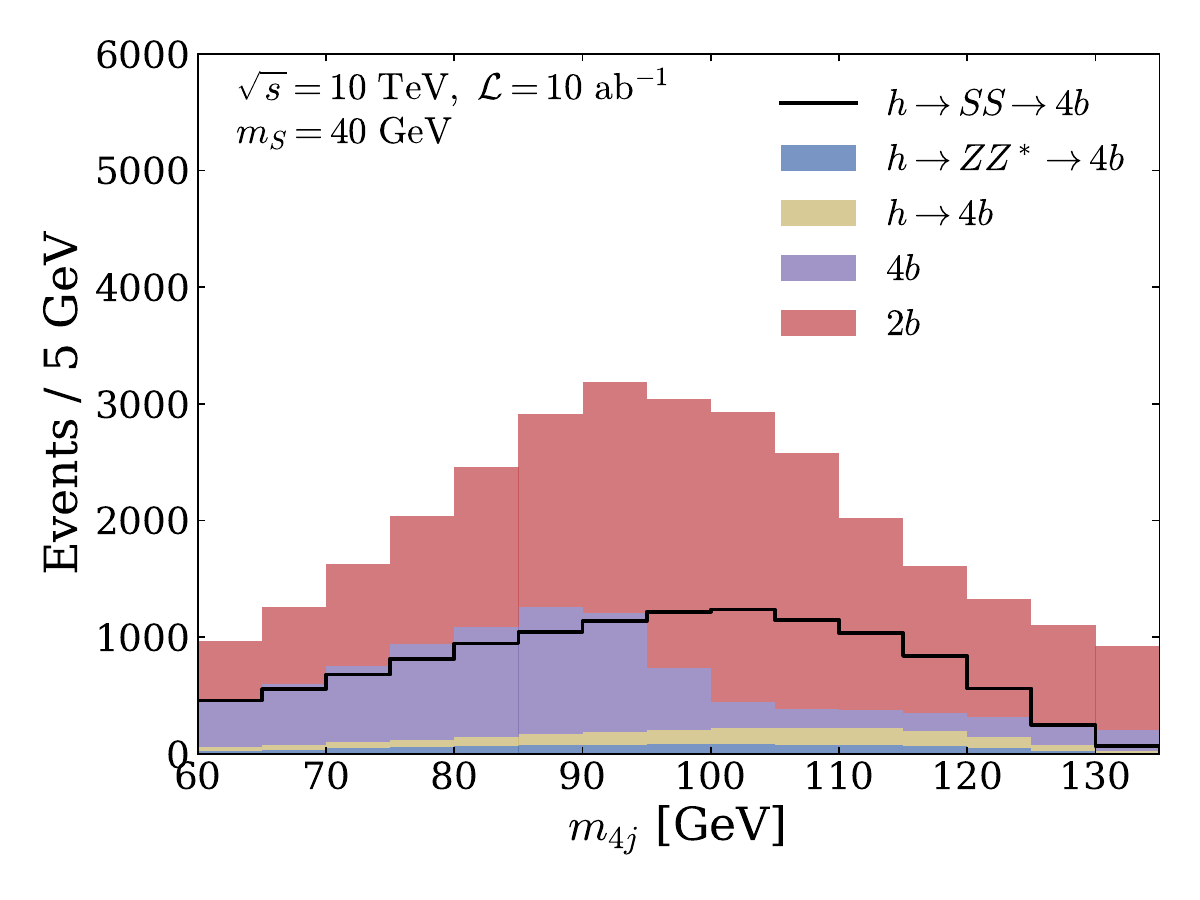}
        \label{fig:m4b_before_dR}
    \end{subfigure}
    \hfill
    \begin{subfigure}[h]{0.48\textwidth}
        \centering
        \includegraphics[width=\textwidth]{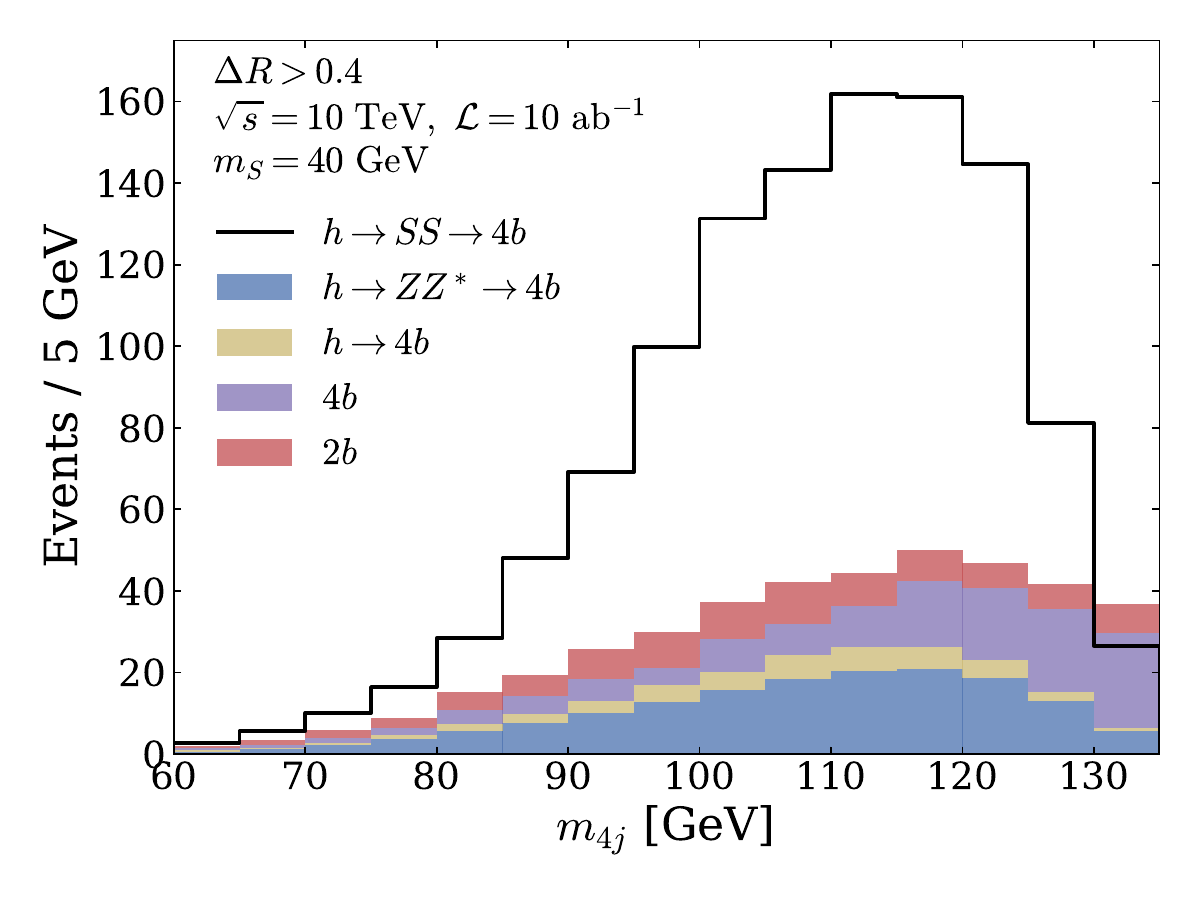}
        \label{fig:m4b_after_dR}
    \end{subfigure}
    \caption{
    Distributions for the invariant mass of the leading four jets in the samples before (left) and after (right) applying the angular separation requirement $\Delta R_{jj} > 0.4$, in the $4b$ channel. The black solid lines indicate the signal distributions while histograms of different colors represent various leading SM backgrounds. The results are for a muon collider operating at $\sqrt{s}=10~\mathrm{TeV}$ with an integrated luminosity of $10~\mathrm{ab}^{-1}$. For the signal, we choose $m_S=$ 40 GeV. We also assume the branching ratio $\mathrm{BR}(h\to SS\to 4b)=10^{-2}$ to make the signal samples similar in size to the background samples after the $\Delta R_{jj}$ cuts. }
\label{fig:m4b_dR_comparison}
\end{figure}

\begin{figure}[htbp]
    \centering
    \begin{subfigure}[h]{0.48\textwidth}
        \centering
        \includegraphics[width=\textwidth]{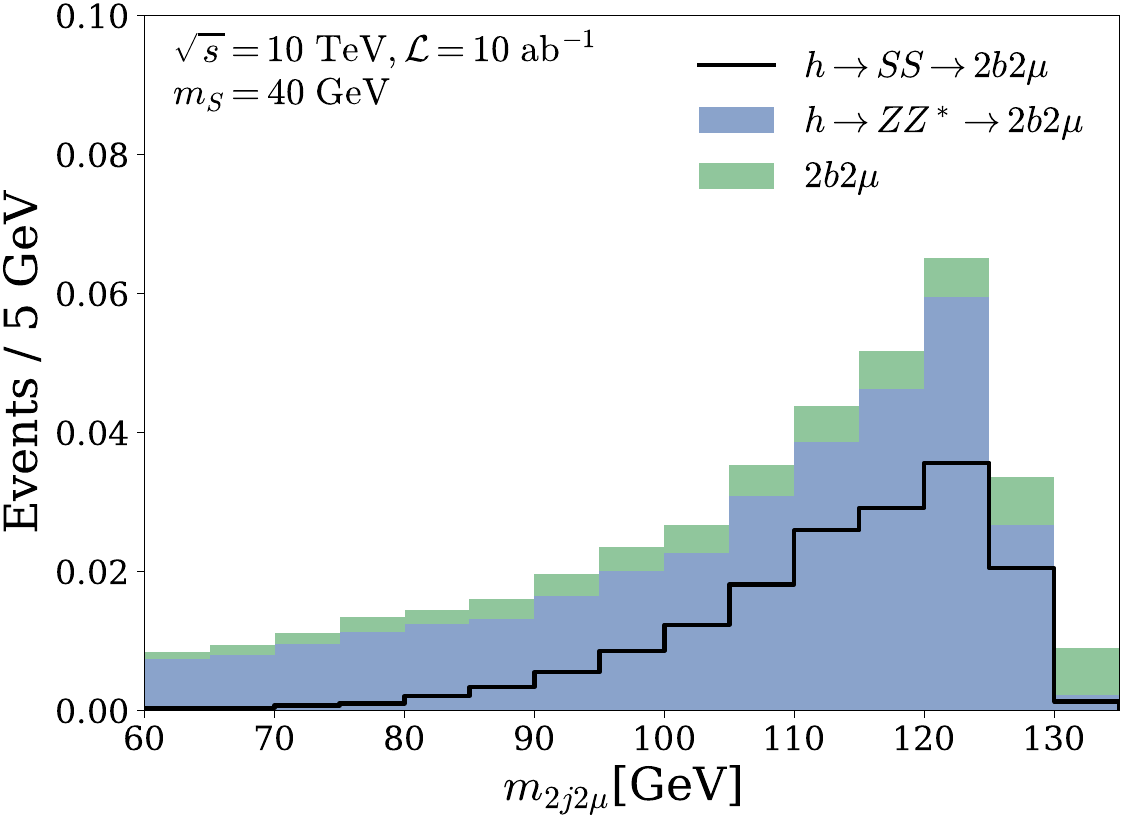}
        \label{fig:m2b2m_before_dR}
    \end{subfigure}
    \hfill
    \begin{subfigure}[h]{0.48\textwidth}
        \centering
        \includegraphics[width=\textwidth]{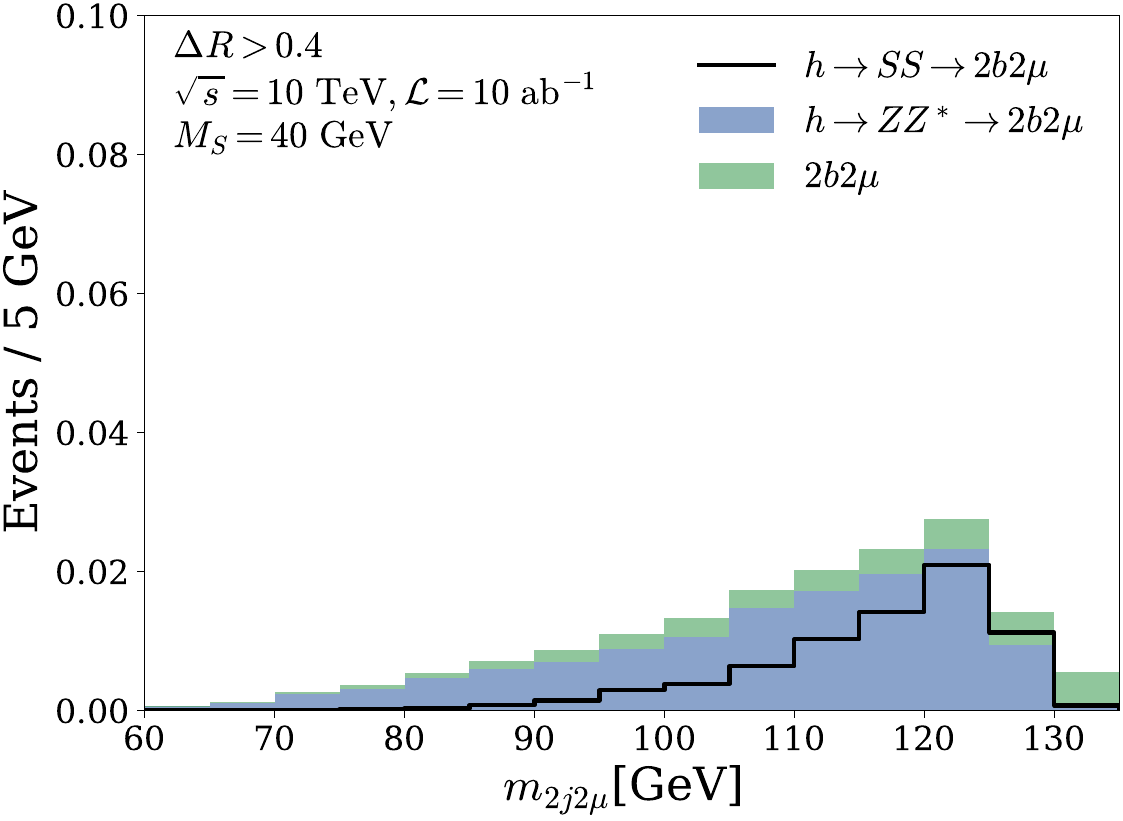}
        \label{fig:m2b2m_after_dR}
    \end{subfigure}
    \caption{
    Invariant mass distributions of the $2b2\mu$ system before (left) and after (right) applying the angular separation requirements $\Delta R_{j\mu,\mu\mu,jj} > 0.4$. The black solid lines indicate the signal distributions while histograms of different colors represent various leading SM backgrounds. The results are for a muon collider operating at $\sqrt{s}=10~\mathrm{TeV}$ with an integrated luminosity of $10~\mathrm{ab}^{-1}$. For the signal, we take the benchmark with $m_S=$ 40 GeV. We also assume a branching ratio $\mathrm{BR}(h\to SS\to 2b2\mu)=10^{-7}$ to make the signal samples similar in size to the background samples after the $\Delta R$ cuts.  }
    \label{fig:m2b2m_dR_comparison}
\end{figure}

\begin{figure}[htbp]
    \centering
    \begin{subfigure}[h]{0.48\textwidth}
        \centering
        \includegraphics[width=\textwidth]{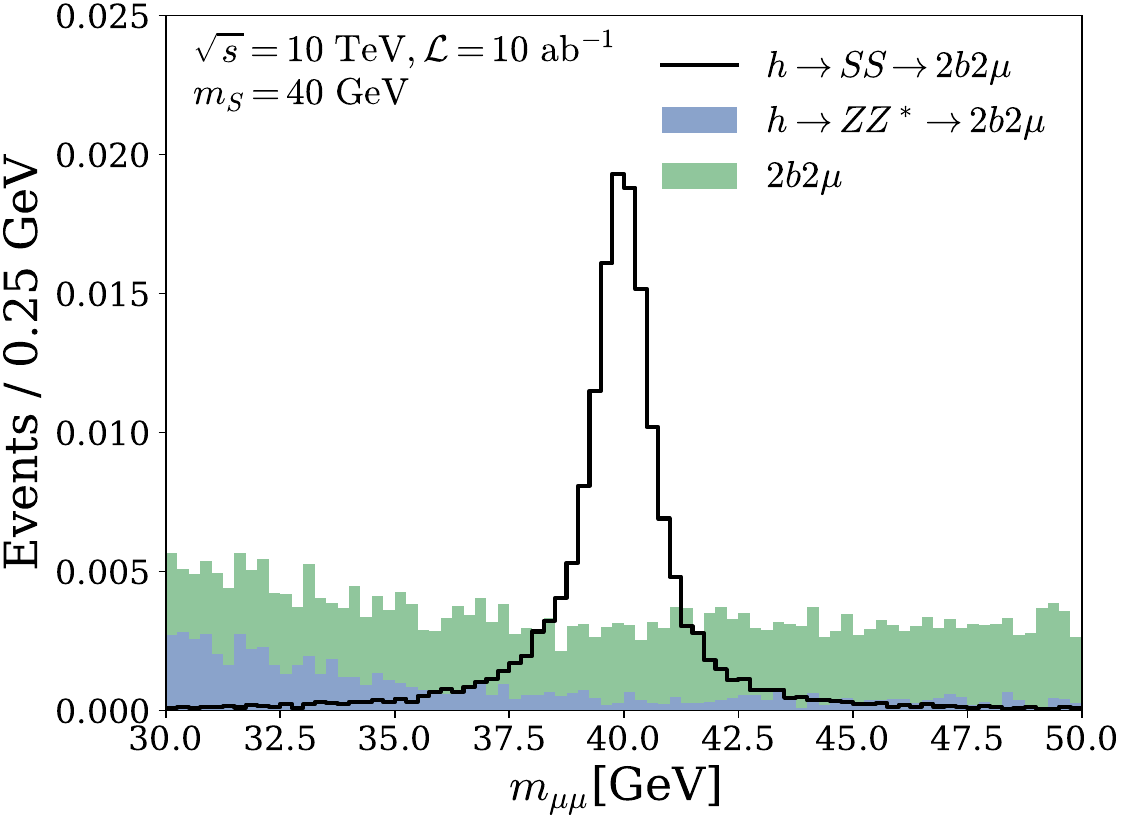}
        \label{fig:m_mumu_before_dR}
    \end{subfigure}
    \hfill
    \begin{subfigure}[h]{0.48\textwidth}
        \centering
        \includegraphics[width=\textwidth]{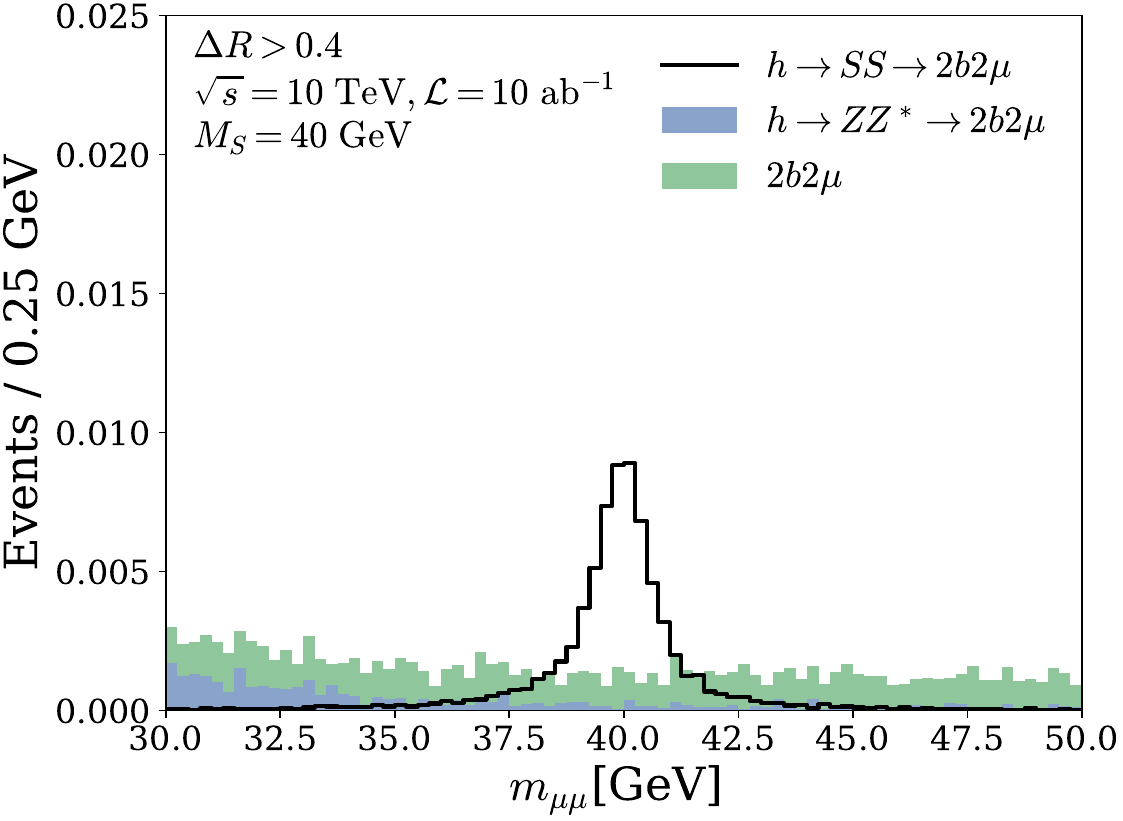}
        \label{fig:m_mumu_after_dR}
    \end{subfigure}
    \caption{
    Invariant mass distributions of muon pairs for the $2b2\mu$ final state before (left) and after (right) applying the angular separation requirements $\Delta R_{jj, j\mu, \mu\mu} > 0.4$. The black solid lines indicate the signal distributions while histograms of different colors represent various leading SM backgrounds. The results are for a muon collider operating at $\sqrt{s}=10~\mathrm{TeV}$ with an integrated luminosity of $10~\mathrm{ab}^{-1}$. For the signal, we choose $m_S=$ 40 GeV. We also assume a branching ratio $\mathrm{BR}(h\to SS\to 2b2\mu)=10^{-7}$. 
   }
    \label{fig:m_mumu_dR_comparison}
\end{figure}

After imposing the $\Delta R_{jj}$ cuts, we show the distributions of the invariant mass of four leading jets, $m_{4j}$, in the right panel Fig.~\ref{fig:m4b_dR_comparison} for the $4b$ final state. For comparison, the distributions before applying the $\Delta R_{jj}$ cut are shown in the left panel. For the signal, the peak of $m_{4j}$ moves from around 100 GeV to (110-120) GeV. This suggests that the $\Delta R_{jj}$ cut suppresses events with overlapping or poorly resolved jets and improves the reconstruction of the intermediate Higgs resonance. The $\Delta R_{jj}$ cut also reduces backgrounds more significantly compared to the signal. On the other hand, the $\Delta R_{jj, j\mu, \mu\mu}$ cuts do not modify shapes of the invariant mass $m_{2b2\mu}$ distributions for the $2b2\mu$ final state as shown in Fig.~\ref{fig:m2b2m_dR_comparison}, as well as those of dimuon invariant mass $m_{\mu\mu}$ distributions in Fig.~\ref{fig:m_mumu_dR_comparison}. We still keep these conventional cuts to be consistent with parton-level cuts.
Given these invariant mass distributions, we impose  further requirements  to select events compatible with an intermediate Higgs resonance: $m_{4b}, m_{2b2\mu} \in [100,150] ~\mathrm{GeV}$ for $4b$ and $2b 2\mu$ final states respectively.
The effects of all the preselection cuts on both the signals and backgrounds are summarized in Table \ref{tab:preselection_4b} and \ref{tab:preselection_2b2m}.

\begin{table}[htbp]
\centering
\footnotesize
\begin{tabular}{lcccc}
\toprule
Process & $\sigma$ [pb] & $p_T^j > 20~\mathrm{GeV}$&$\Delta R_{jj} > 0.4$&$100~\mathrm{GeV} < m_{4j} < 150~\mathrm{GeV}$ \\
\midrule
\multicolumn{5}{l}{\textbf{Signal}} \\

$h \to SS \to 4b$ 
& $0.84 \times \mathrm{BR}$ 
& $1.6 \times 10 ^{-1}$ & $1.3 \times 10 ^{-2}$ & $1.0 \times 10 ^{-2}$ \\

\midrule
\multicolumn{5}{l}{\textbf{Background}} \\

$h \to ZZ^* \to 4b$ 
& $5.0 \times 10^{-4}$ 
& $1.9 \times 10 ^{-1}$ & $3.2 \times 10 ^{-2}$ & $2.3 \times 10 ^{-2}$  \\


$h \to 4b$ 
& $1.0 \times 10^{-3}$ 
& $1.5 \times 10 ^{-1}$ & $4.3 \times 10 ^{-3}$ & $2.9 \times 10 ^{-3}$ \\

$2b$ 
& $2.0 \times 10^{-2}$ 
& $3.0 \times 10 ^{-1}$ & $1.6 \times 10 ^{-3}$ & $8.4 \times 10 ^{-4}$\\

$4b$ 
& $8.7 \times 10^{-3}$ 
& $7.4 \times 10 ^{-1}$ & $9.2 \times 10 ^{-2}$ & $1.1 \times 10 ^{-3}$ \\

\bottomrule
\end{tabular}
\caption{
Preselection cutflow table for both the signal and background processes in the $4b$ final state at a 10 TeV muon collider. We list all cross sections. The cross section of the signal is computed as the Higgs production cross section times the branching ratio of exotic Higgs decay. BR in this table stands for the branching ratio BR$(h \to SS \to 4b)$. After each cut, we list the remaining fraction of events. Here we only consider the dominant $W$-boson fusion processes. The neutral-boson fusion contributions are sub-dominant and only give small corrections to the final results. 
}
\label{tab:preselection_4b}
\end{table}

\begin{table}[htbp]
\centering
\footnotesize
\begin{tabular}{lcccc}
\toprule
Process & $\sigma$ [pb] & $p_T^j > 20~\mathrm{GeV}$&$\Delta R_{jj, j\mu, \mu\mu} > 0.4$& 100 GeV $<m_{2j2\mu}<$150 GeV \\
& &  $p_T^\mu > 5~\mathrm{GeV}$ & & \\
\midrule
\multicolumn{5}{l}{\textbf{Signal}} \\

$h \to SS \to 2b2\mu$ 
& $0.84 \times \mathrm{BR}$ 
 &$2.0 \times 10 ^{-1}$ & $8.6 \times 10 ^{-2}$&$8.0 \times 10 ^{-2}$   \\

\midrule
\multicolumn{5}{l}{\textbf{Background}} \\

$h \to ZZ^* \to 2b2\mu$ 
& $1.8 \times 10^{-3}$ 
& $7.7 \times 10 ^{-2}$ & $2.9 \times 10 ^{-2}$ & $2.4 \times 10 ^{-2}$  \\

$2b2\mu$ 
& $2.0 \times 10^{-2}$ 
& $7.4 \times 10 ^{-2}$ & $3.0 \times 10 ^{-2}$& $3.0 \times 10 ^{-3}$ \\

\bottomrule
\end{tabular}
\caption{
Preselection cutflow table for signal and background processes in the $2b2\mu$ final state at a 10 TeV muon collider. BR in this table stands for the branching ratio BR$(h \to SS \to 2b2\mu)$. Similar to Table~\ref{tab:preselection_4b}, we list all cross sections and fractions of surviving events after each step. We only consider the dominant $W$-boson fusion processes. 
}
\label{tab:preselection_2b2m}
\end{table}

\subsection{Machine Learning Selection}
For event selection, especially in the 4$b$ channel, due to QCD radiation and jet combinatorics, traditional cut-based methods often struggle on background mitigation and fail to capture intricate correlations between kinematic variables such as invariant masses, $\Delta R$, and other dynamic characteristics of final-state particles. Thus, we apply machine learning (ML) techniques to form a binary classifier to improve the analysis after imposing the preselection cuts above. We use the Boosted Decision Tree (BDT) based ML algorithm XGBoost~\cite{Chen:2016btl}, also known as Extreme Gradient Boosting. It is a widely adopted algorithm in particle physics due to its efficiency, scalability, and superior performance in handling high-dimensional datasets typical of high-energy physics. The algorithm builds a BDT ensemble through gradient boosting with several optimizations and assembles them in a sequential boosting ensemble, where each new tree fits the residuals of the current model using second-order gradients and Hessians for precise split selection. 

In our analysis, we apply \texttt{XGBoost 2.1.4} after the preselection cuts. Before training, background samples are reweighted according to their expected yields after preselection in Table~\ref{tab:preselection_4b} and~\ref{tab:preselection_2b2m}. We then randomly divide sample events into training sets and test sets with their sizes shown in Table~\ref{tab:TrainTest}. The input parameters for the $4b$ final state are $(p_T,\eta,\phi)$ of each jet and possible 6 jet pairs, invariant mass of each jet pair $m_{jj}$ and its corresponding difference to the chosen $m_S$ benchmark ($|m_{jj}-m_S|$). For the $2b2\mu$ final state, the inputs are prepared in a similar manner. However, since the signal to background ratio is highly sensitive to $m_{\mu\mu}$, relevant information will be excluded from the input to improve the overall performance. In this channel, the inputs include $(p_T,\eta,\phi)$ for all individual objects and the jet/muon pairs aside from those vetoed variables. Transverse momenta $p_T$'s of each muon and muon pair are vetoed, leaving the $m_{\mu\mu}$ information inaccessible to the ML model. We also include the invariant mass of the jet pair $m_{jj}$, and its difference to the chosen $m_S$ benchmark $|m_{jj}-m_S|$ as in the $4b$ channel. 

\begin{table}[htbp]
\centering
\footnotesize
\begin{tabular}{lcc}
\toprule
Process & Training set size & Testing set size  \\
\midrule
\multicolumn{3}{l}{\textbf{Signal}} \\

$h \to SS \to 4b$ 
& 46280
& 46280\\
$h \to SS \to 2b2\mu$ 
& 4337
& 4337\\

\midrule
\multicolumn{3}{l}{\textbf{Background}} \\

$h \to ZZ^* \to 4b$ 
& 277581
& 277581 
 \\


$h \to 4b$ 
& 695265
& 695265 
 \\

$2b$ 
& 800000 
& 800000
\\

$ 4b$ 
& 596704
& 596704
 \\
\hline
$h\to Z Z^{*} \to 2b2\mu$
& 1622
& 1622
 \\

$2b 2\mu$
& 3962
& 3962
 \\

\bottomrule
\end{tabular}
\caption{Training and testing set sizes for both $4b$ and $2b2\mu$ final states. The first four backgrounds are for $4b$ final state while the last two are for $2b2\mu$ final state. The training and testing set sizes are forced to be the same for each $m_S$ benchmark simulated. }
\label{tab:TrainTest}
\end{table}

To obtain stable performance, we apply batch normalization and train 5 parallel models, each with different initialization and hyperparameter tuning. The final BDT output for selection, namely the BDT score, is the average output of the five. 
For the $4b$ channel, the features' contribution to signal-background discrimination are ranked after training. Sorted by importance, the most important feature is the invariant mass of the jet pair $m_{jj}$ with the smallest deviation from the chosen $m_S$ benchmark, followed by the minimum invariant mass difference between two jet pairs in each event, the minimum mass difference to the $m_S$ benchmark $|m_{jj}-m_S|$, and $\Delta R_{jj}$. Though $m_{jj}$ with the smallest deviation from the $m_S$ benchmark is the dominant one, other features still contribute significantly. 
For the training of the final state $2b2\mu$, the leading feature for discriminating background is the invariant mass of the jet pair $m_{jj}$, while all other features are much less effective.

\begin{figure}[htbp]
    \centering
    \includegraphics[width=0.8\textwidth]{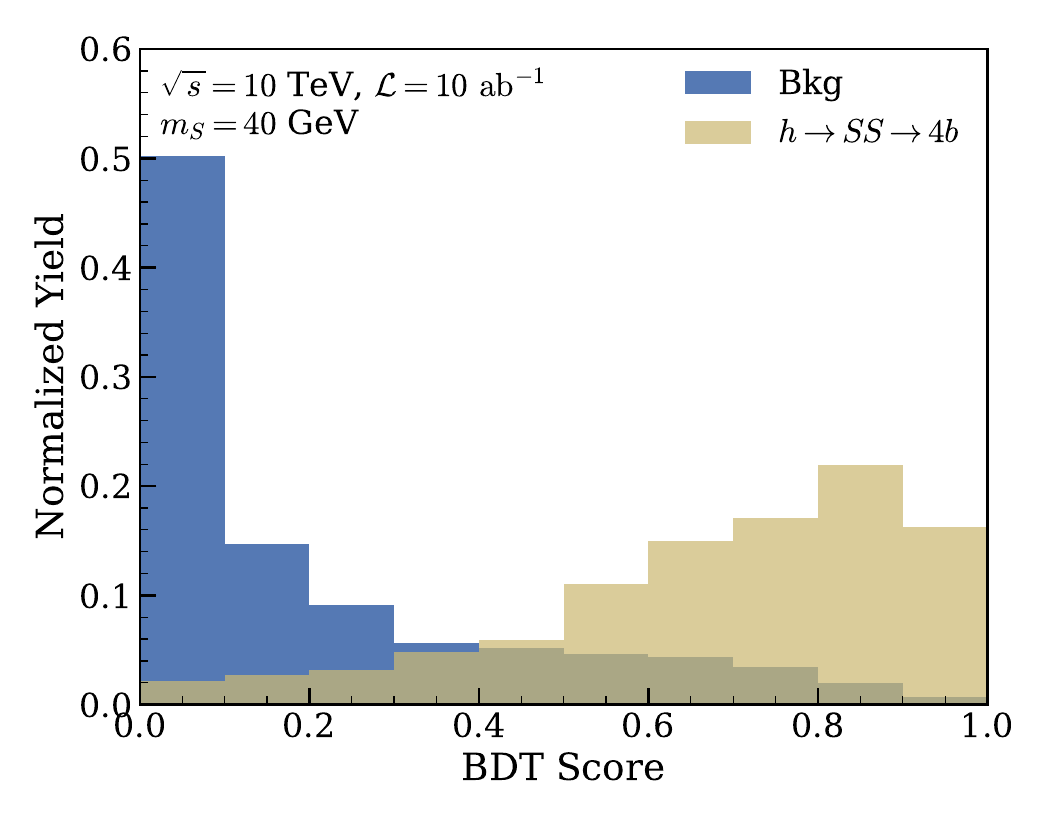}
    \caption{Normalized BDT score distributions for the $4b$ channel. The background distribution (blue) distinguishes clearly from the signal one (brown). The BDT score ranges from 0 to 1 since we apply averaged ensemble model here which involves a sigmoid function in the output. The yield is computed for a 10 TeV muon collider with 10 ab$^{-1}$ data. }
    \label{fig:BDT}
\end{figure}

\begin{figure}[htbp]
    \centering
    \begin{subfigure}[t]{0.48\textwidth}
        \centering
        \includegraphics[width=\textwidth]{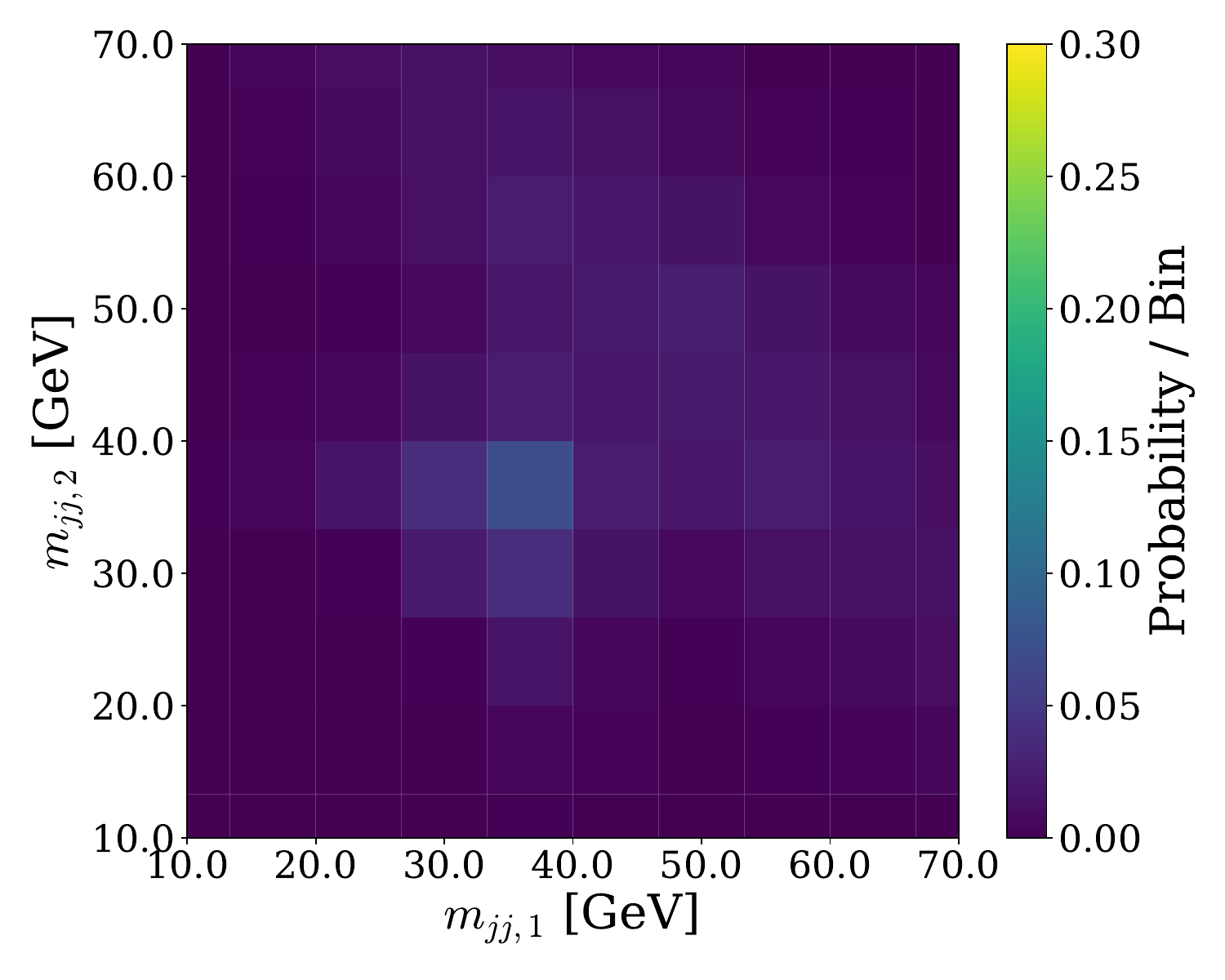}
    \end{subfigure}
    \hfill
    \begin{subfigure}[t]{0.48\textwidth}
        \centering
        \includegraphics[width=\textwidth]{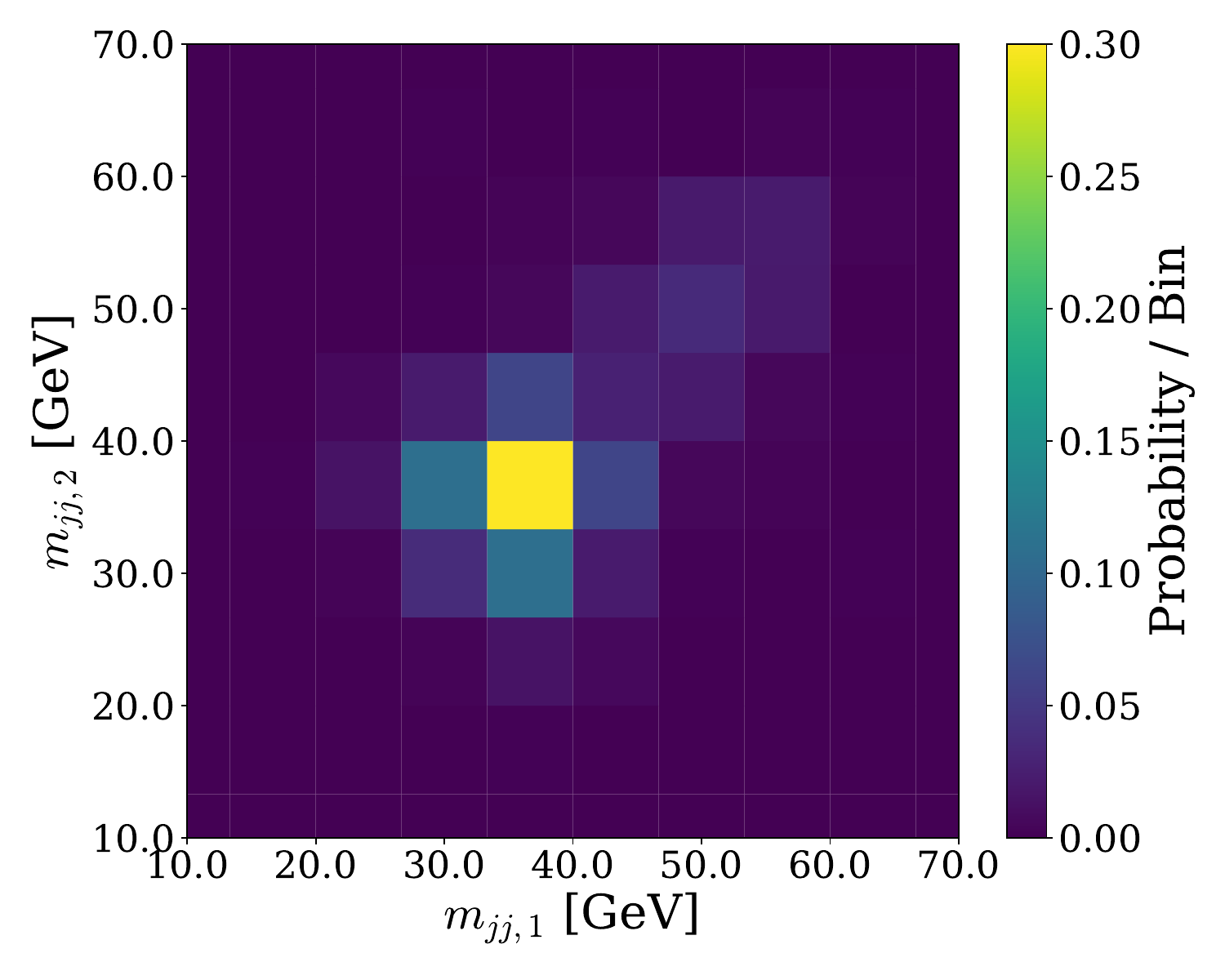}
    \end{subfigure}
    \caption{Normalized signal distributions in the plane of invariant masses of two jet pairs in the $4b$ channel. The left panel is after the preselection cuts but before the ML cuts while the right one is after the ML cuts but before the $b$-tagging cut. Both jet pairs' invariant masses concentrate closer to the benchmark $m_S=40$~GeV after ML selection. } 
    \label{fig:signal distribution}
\end{figure}

After training, one could obtain the distribution of the BDT output for each $m_S$ benchmark. The signal and background regions form two separate peaks, as shown in Fig~\ref{fig:BDT}. For each $m_S$ benchmark, we calculate the ${S/\sqrt{B}}$ value for every threshold BDT score value, and choose the threshold value with the maximum ${S/\sqrt{B}}$ value to be the BDT score of the benchmark that will be used in ML selection. After applying such threshold value to ML selection, we find that in terms of ML Area-Under-Curve (AUC) distributions,\footnote{The closer the AUC value is to 1, the better the model fits.} the values for the $2b2\mu$ final state are in the range $[0.92, 0.99]$, which are evidently higher than the corresponding values of the $4b$ final state in the range $[0.85, 0.91]$. All these AUC values are close to 1, indicating that ML models are indeed able to distinguish signals from backgrounds effectively. The lower AUC values for the $4b$ final state originates from the fact that the QCD radiation and combinatorics of $4b$ make it more difficult to select signal events out of backgrounds. 
For the $4b$ final state, distributions in the jet-pair invariant mass plane are demonstrated in Fig.~\ref{fig:signal distribution}, before and after the ML cuts. The $x$ and $y$ axes are the invariant masses of two jet pairs in each event. It is obvious that after ML selection, both invariant masses shift closer to $m_S$, demonstrating that ML algorithms work to identify the right jet pairing.

 After applying the ML cuts, we further impose a few more cuts to improve the sensitivity. For the $4b$ final state, we require that there should be 4 $b$-tagged jets in each event. For the $2b 2\mu$ final state, we first require that there should be 2 $b$-tagged jets in the event. Then we choose a proper $m_{\mu\mu}$ mass window which optimizes the sensitivity for each mass benchmark. We find that the signal and background efficiencies of these selection rules are about the same when applied to all the samples before and after the ML procedure. This fact indicates a low correlation between inputs to the ML model and $b$-tagging or $m_{\mu\mu}$ values. Including them afterward as independent selection criteria makes it easy to generate large samples for ML training and avoids the ML classifier being dominated by these quantities. The final cut-flow tables for the $m_S=40$ GeV benchmark are given in Table~\ref{tab:cutflow_4b_ml} and~\ref{tab:cutflow_2b2m_ml}. The final yields for all $m_S$ benchmarks are given in the appendix.

\begin{table}[htbp]
\centering
\footnotesize
\begin{tabular}{lccccc}
\toprule
Process & $\sigma$ [pb] & Preselection & ML selection & $4b$-tagging & Yield \\
\midrule
\multicolumn{6}{l}{\textbf{Signal}} \\

$h \to SS \to 4b$ 
& $0.84 \times \mathrm{BR}$ 
& $1.0 \times 10^{-2}$ 
& $7.0 \times 10^{-3}$ 
& $1.9 \times 10^{-3}$ 
& $1.3 \times 10^{4} \times \mathrm{BR}$ \\

\midrule
\multicolumn{6}{l}{\textbf{Background}} \\

$h \to ZZ^* \to 4b$ 
& $5.0 \times 10^{-4}$ 
& $2.3 \times 10^{-2}$ 
& $6.6 \times 10^{-3}$ 
& $1.4 \times 10^{-3}$ 
& $7.0$ \\

$h \to 4b$ 
& $1.0 \times 10^{-3}$ 
& $2.9 \times 10^{-3}$ 
& $1.3 \times 10^{-3}$ 
& $2.9 \times 10^{-4}$ 
& $2.9$ \\

$2b$ 
& $2.0 \times 10^{-2}$ 
& $8.4 \times 10^{-4}$ 
& $2.7 \times 10^{-4}$ 
& $\leq 3.8 \times 10^{-6}$ 
& $\leq 0.76$ \\

$4b$ 
& $8.7 \times 10^{-3}$ 
& $1.1 \times 10^{-3}$ 
& $2.5 \times 10^{-4}$ 
& $1.7 \times 10^{-5}$ 
& $1.5$ \\

\bottomrule
\end{tabular}
\caption{
Cutflow table for signal and background processes in the $4b$ final state at a 10 TeV muon collider with an integrated luminosity of $10~\mathrm{ab}^{-1}$. We choose $m_S = 40$ GeV. 
The preselection cuts are described in Sec.~\ref{sec:preselection}. For the signal, BR stands for branching ratio BR$(h \to SS \to 4b)$. Similar to Table~\ref{tab:preselection_4b}, we list all cross sections and the fraction of remaining events after each set of cuts. We also provide the final yields (the number of events after all the cuts). 
}
\label{tab:cutflow_4b_ml}
\end{table}

\begin{table}[htbp]
\centering
\footnotesize
\begin{tabular}{lcccccc}
\toprule
Process & $\sigma$ [pb] & Preselection & ML selection & $2b$-tagging & $m_{\mu\mu}$& Yield \\
\midrule
\multicolumn{7}{l}{\textbf{Signal}} \\

$h \to SS \to 2b2\mu$ 
&0.84$\times \mathrm{Br}$& $8.0 \times 10 ^{-2}$ & $ 7.8\times 10 ^{-2}$ & $ 3.8\times 10 ^{-2}$ &$3.7\times 10^{-2}$ & $3.1\times 10^5 \times$Br \\

\midrule
\multicolumn{7}{l}{\textbf{Background}} \\

$ h\!\to  2b2\mu$  &
$ 4.4 \times 10 ^{-4}$ &  $2.4 \times 10 ^{-2}$ & $4.2 \times 10 ^{-3}$ & $ 1.4\times 10 ^{-3}$ & $2.2\times 10^{-4}$ &  0.96\\

$2b2\mu$  &
$ 1.2 \times 10 ^{-3}$ &  $ 3.0\times 10 ^{-3}$ & $9.8 \times 10 ^{-4}$ & $ 3.9\times 10 ^{-4}$ & $1.0\times 10^{-4}$ & 1.2\\

\bottomrule
\end{tabular}
\caption{
Cutflow table for signal and background processes in the $2b2\mu$ final state at a 10 TeV muon collider with an integrated luminosity of $10~\mathrm{ab}^{-1}$. We choose $m_S = 40$ GeV. The preselection cuts are described in Sec.~\ref{sec:preselection}. BR in this table represents BR$(h \to SS \to 2b2\mu)$. 
We add a $m_{\mu\mu}$ cut after ML cuts, where we apply a $m_{\mu\mu}$ invariant mass window with width in the range $[0,10]$ GeV and calculate the corresponding sensitivity. We choose the ${m_{\mu\mu}}$ mass window that gives us the best sensitivity as the final cut. We list all cross sections, the fraction of remaining events after each set of cuts, and final yields.  }
\label{tab:cutflow_2b2m_ml}
\end{table}

\subsection{Results}

With the signal and background efficiencies, we could compute minimum branching ratios of exotic Higgs decays for different $m_S$'s that a muon collider could probe. We estimate the signal significance using $S/\sqrt{S+B+\delta B^2}$ with $S(B)$ denoting the signal (background) counts in the signal region. We take $\delta B= 0.05 B$, presuming the background systematic uncertainty is of $5\%$. Then the minimum branching ratio that a muon collider is sensitive to is obtained by setting $S/\sqrt{S+B+\delta B^2}=1.96$, corresponding to a $95\%$ confidence level (CL).

\begin{figure}[htbp]
    \centering
    \begin{subfigure}[t]{0.48\textwidth}
        \centering
        \includegraphics[width=\textwidth]{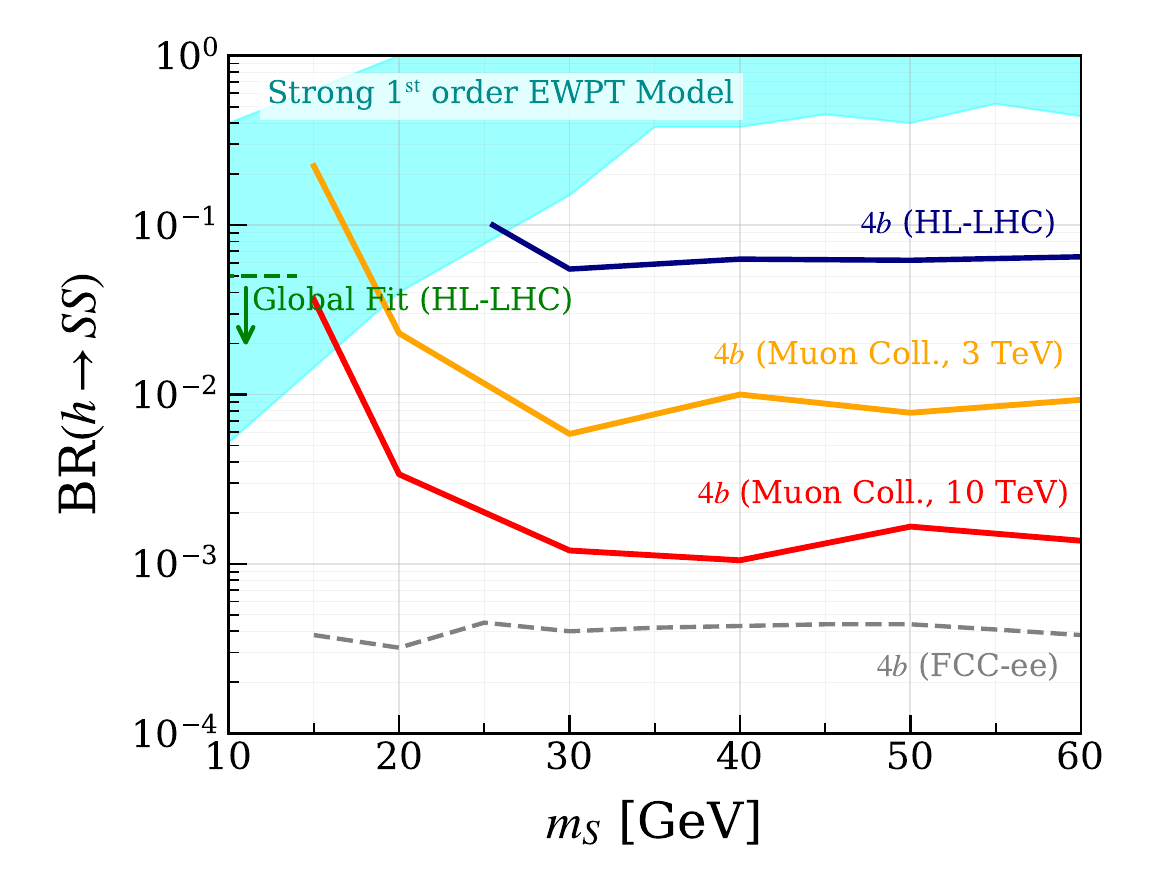}
    \end{subfigure}
    \hfill
    \begin{subfigure}[t]{0.48\textwidth}
        \centering
\includegraphics[width=\textwidth]{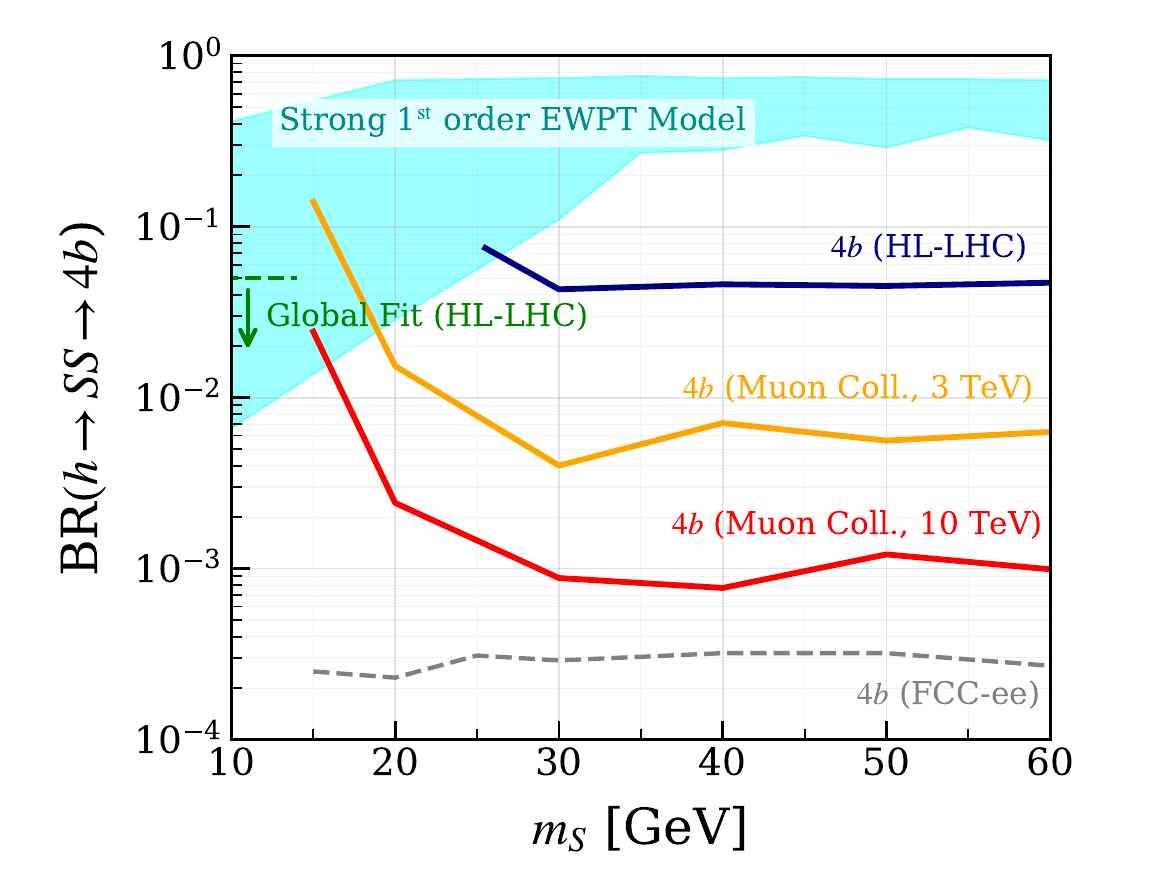}
    \end{subfigure}
    \caption{Projected sensitivities to ${\rm BR} (h \to SS)$ in the Higgs-singlet mixing model from analysis of the $4b$ channel (left) and ${\rm BR} (h \to SS \to 4b)$ without specifying how $S$ decays (right). The red solid curves show projected 95\% CL limits at a 10 TeV muon collider with 10 ab$^{-1}$ data while the yellow solid curves show projected 95\% CL limits at a 3 TeV muon collider with 1 ab$^{-1}$ data. For comparison, we also show projected 95\% CL limits at HL-LHC as blue solid curves (rescaled based on~\cite{ATLAS:2018pvw}) and at a Higgs factory like FCC-$ee$ (rescaled based on~\cite{Wang:2023zys}) as gray dashed lines. In addition, projected HL-LHC global-fit upper limit on the inclusive branching ratio of
Higgs decays beyond the SM \cite{cepeda:2019klc} is shown as the horizontal green dashed lines and the parameter space compatible with a strong first-order EWPT is shown as blue-shaded regions~\cite{Kozaczuk:2019pet}. }
    \label{fig:BR for 4b}
\end{figure}

\begin{figure}[htbp]
    \centering
    \begin{subfigure}[t]{0.48\textwidth}
        \centering
        \includegraphics[width=\textwidth]{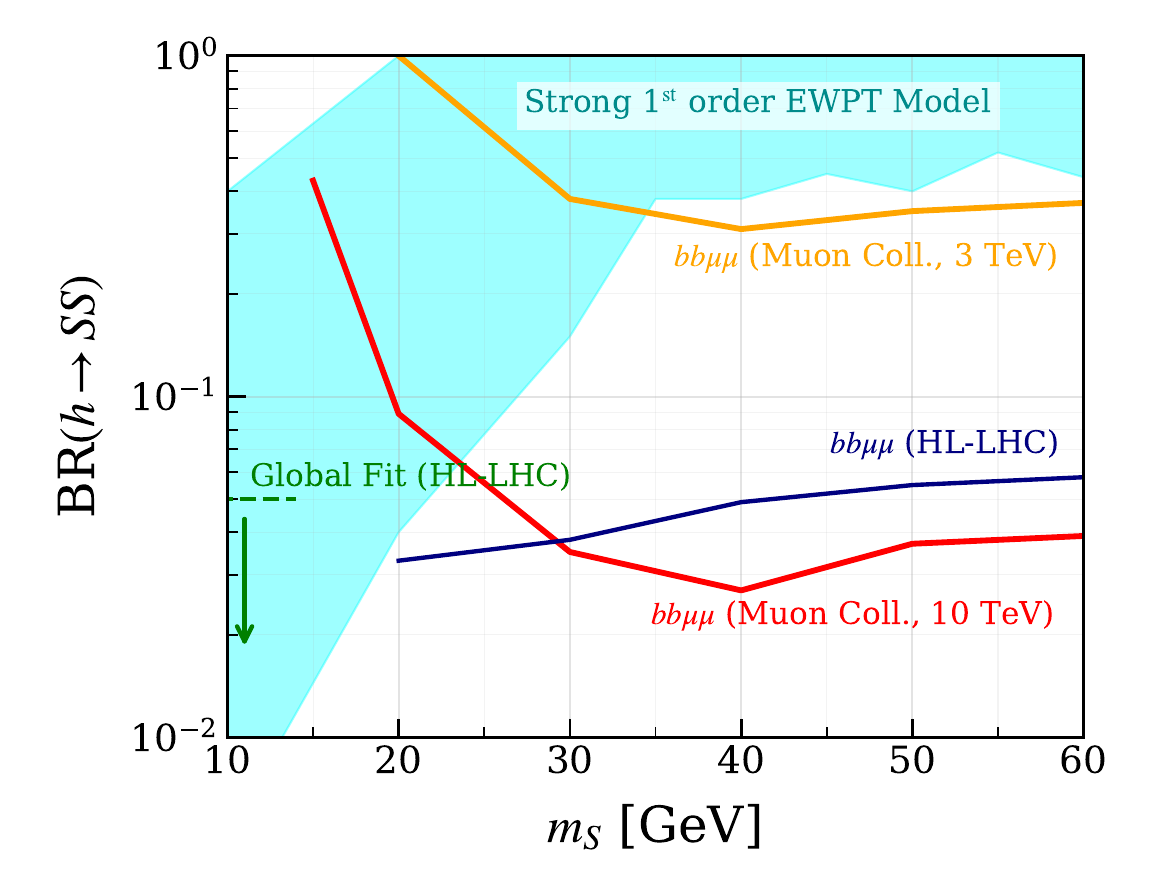}
    \end{subfigure}
    \hfill
    \begin{subfigure}[t]{0.48\textwidth}
        \centering
        \includegraphics[width=\textwidth]{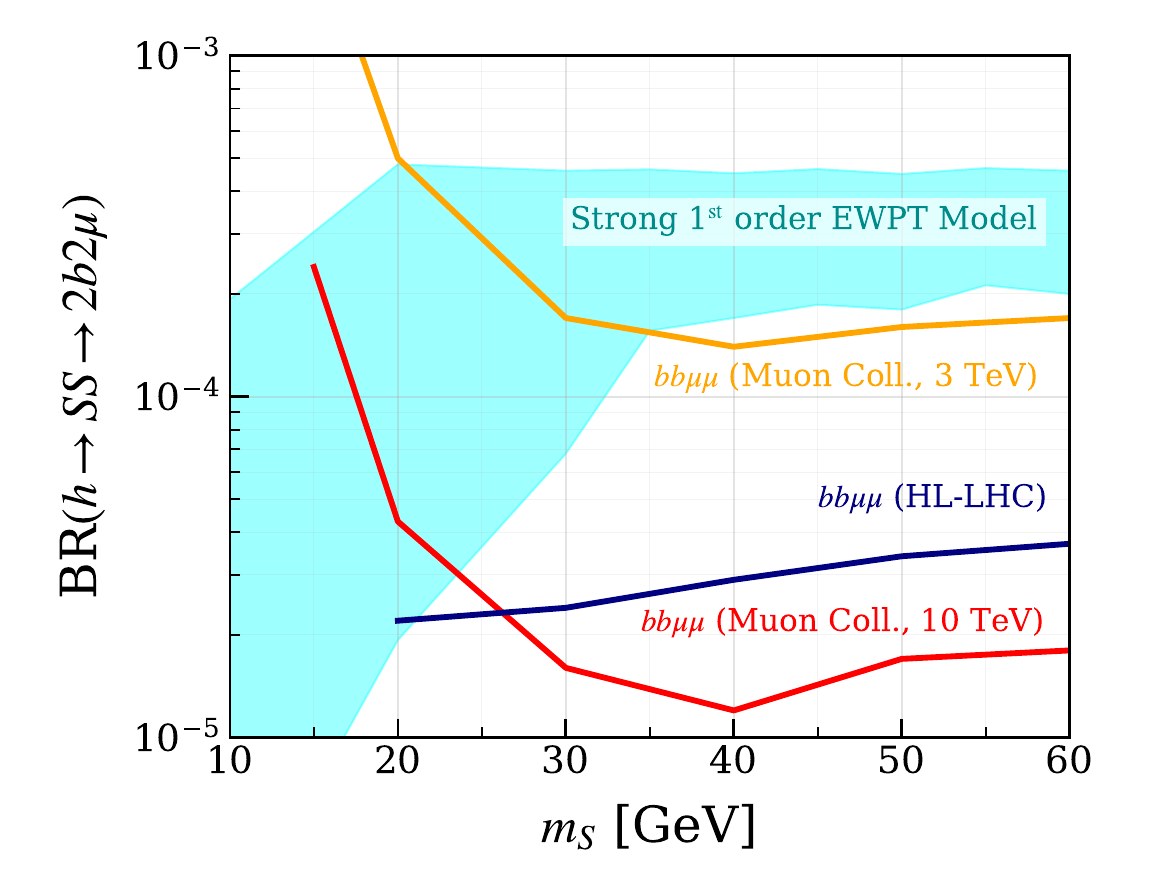}
    \end{subfigure}
    \caption{Projected sensitivities to ${\rm BR} (h \to SS)$ in the Higgs-singlet mixing model from analysis of the $2b 2\mu$ channel (left) and ${\rm BR} (h \to SS \to 2b2\mu)$ without specifying how $S$ decays (right). The red solid curves show projected 95\% CL limits at a 10 TeV muon collider with 10 ab$^{-1}$ data while the yellow solid curves show projected 95\% CL limits at a 3 TeV muon collider with 1 ab$^{-1}$ data. For comparison, we also show projected 95\% CL limits at HL-LHC as blue solid curves (rescaled based on \cite{ATLAS:2018emt, CMS:2018nsh}). In addition, projected HL-LHC global-fit upper limit on inclusive
exotic Higgs decays~\cite{cepeda:2019klc} is shown as the horizontal green dashed line and the parameter space compatible with a strong first-order EWPT is shown as blue-shaded regions~\cite{Kozaczuk:2019pet}. }
    \label{fig:BR for 2b2m}
\end{figure}


The final results for a 3 TeV or a 10 TeV muon collider (with integrated luminosities of of $1$ or $10~\mathrm{ab}^{-1}$ respectively) are shown in Fig.~\ref{fig:BR for 4b} and Fig.~\ref{fig:BR for 2b2m}. The left panel of Fig.~\ref{fig:BR for 4b} shows the projected 95\% CL limits on the branching ratio $\mathrm{BR}(h \to SS)$ as a function of $m_S$ in the benchmark model described in Sec.~\ref{sec:models and simulations} with $S$-decay branching ratios entirely determined by the scalar mass. We also show the general projected 95\% CL limits on $\mathrm{BR}(h \to SS \to 4b)$ without specifying decay branching ratios of $S$ in the right panel of Fig.~\ref{fig:BR for 4b}. 
From the figure, we could see that a 10 TeV muon collider with $10~\mathrm{ab}^{-1}$ data can probe branching ratio $\mathrm{BR}(h \to SS)$ at the level of $\mathcal{O}(10^{-3})$ for $m_S > 20$ GeV, surpassing the projected HL-LHC reach by almost two orders of magnitude.
The improvement is particularly pronounced for $m_S \sim (30$--$40)~\mathrm{GeV}$, where backgrounds are efficiently suppressed and the Higgs mass reconstruction is most effective. Conversely, for a light $S$ with $m_S \lesssim 20$~GeV, the sensitivity drops significantly. In the low mass region, the more collimated $b$-jet pairs from light $S$ decays have lower chances to produce four resolved jets. The probability for a signal event to pass the minimum $\Delta R$ cut is also lower, leading to much weakened limits for $m_S \lesssim 20$ GeV. Since the singlet scalar $S$ predominantly decays into bottom quarks over a wide mass range, the $4b$ channel benefits from the largest signal rate in the benchmark model described in Sec.~\ref{sec:models and simulations}. This is also the reason that the limits on $\mathrm{BR}(h \to SS)$ in the Higgs-singlet mixing model are similar to the general limits on $\mathrm{BR}(h \to SS \to 4b)$ without specifying $\mathrm{BR}(S \to b\bar{b})$. Limits for the 3~TeV scenario are shown in both plots as yellow curves, which have analogous behavior as their 10~TeV counterparts but are weaker by less than one order of magnitude due to the smaller luminosity and Higgs production rate. The overall limits are of $\mathcal{O}(10^{-2})$ level for $m_S \gtrsim 20$ GeV. For comparison, we also show the projected HL-LHC global-fit upper limit on inclusive exotic Higgs decays~\cite{cepeda:2019klc} in Fig.~\ref{fig:BR for 4b} as the horizontal dashed line in each panel, which is exceeded by both 3 and 10 TeV muon-collider runs when $m_S>20$~GeV. The parameter space compatible with a strong first-order EWPT~\cite{Kozaczuk:2019pet} is presented as blue shaded areas in both plots. Except for the small $m_S$ region, both 3 and 10 TeV running can probe this region well. For completeness, we also include the limit from future Higgs factories such as FCC-$ee$ or CEPC~\cite{FCC:2025lpp,FCC:2025uan,Ai:2025cpj} for the $4b$ channel in Fig.~\ref{fig:BR for 4b}, assuming an integrated luminosity of 5~ab$^{-1}$~\cite{Wang:2022dkz}. As indicated, a Higgs factory is more capable of measuring exotic Higgs decays. Compared to the 10~TeV muon collider benchmark, its overall Higgs yield is $\mathcal{O}(10)$ times smaller. However, the signal efficiency and signal to background ratio at a Higgs factory benefit strongly from the low background level and excellent global energy conservation at $\sqrt{s}=240$~GeV, resulting in strong projected limits in the $4b$ channel. Nevertheless, the muon-collider reach could be further improved if the stringent preselection rules are relaxed, which would reduce the associated loss in the signal efficiency and bring the sensitivities closer to the Higgs-factory ones. In that case, a more sophisticated machine-learning-based strategy (see, e.g., Refs.~\cite{Hajer:2015gka,Craig:2016ygr,Li:2019wpa,Matheus:2026dmj}) will be needed to mitigate backgrounds. Such an approach may also help recover sensitivity in the low $m_S$ region.

The left panel of Fig.~\ref{fig:BR for 2b2m} shows the projected 95\% CL limits on the branching ratio $\mathrm{BR}(h \to SS)$ as a function of $m_S$ in the Higgs-singlet mixing model, from the analysis of $2b2\mu$ final state. The right panel shows the general projected 95\% CL limits on $\mathrm{BR}(h \to SS \to 2b2\mu)$ without specifying $S$'s decay branching ratio into $2b2\mu$. 
This channel offers a much cleaner experimental signature due to the presence of a dimuon pair.
In a model-independent framework without specifying how $S$ decays, the $2b2\mu$ channel exhibits an excellent sensitivity: a 10 TeV muon collider with 10 ab$^{-1}$ data could probe $\mathrm{BR} (h \to SS \to 2b2\mu)$ close to $10^{-5}$, about a factor of (2-4) improved over the reach of HL-LHC for $m_S > 30$ GeV. This suggests a strong background suppression achievable with precise dimuon resonance reconstruction. However, in the specific Higgs-portal scenario considered here, the sensitivity to the exotic Higgs decays in the $2b2\mu$ channel is intrinsically limited by the small branching ratio of $S$ decaying into muons $\mathrm{BR}(S \to \mu^+\mu^-)$, as shown in the left panel of Fig.~\ref{fig:BR for 2b2m}. 
As a result, the $4b$ channel would be the primary discovery one for the exotic decays $h \to SS$ in the Higgs portal model while the $2b2\mu$ channel has a much weaker reach. Similar to the $4b$ channel, the 3~TeV scenario bounds shown as yellow curves are about one order of magnitude weaker than their 10~TeV counterparts.

\section{Conclusions}
\label{sec:con}

A muon collider is commonly envisioned as a powerful facility for precision studies of EWSB and for new physics searches. As an effective high-energy electroweak boson collider, it combines a sizable Higgs production rate mainly through VBF with a substantially cleaner environment than hadron colliders. It is therefore well suited to probe Higgs boson's interactions with (partially) hadronic final states and small rates of related exotic processes. In this work, we focus on exotic Higgs decays induced by Higgs mixing with a beyond-SM singlet. The scenario naturally arises in Higgs-portal new physics and can be closely connected to questions such as the structure of scalar sector in the SM and beyond as well as the nature of EWPT. Two decay chains, namely $h\to SS\to 4b$ and $h\to SS\to 2b2\mu$ are studied, at two muon collider operation scenarios with $\sqrt{s}$ = 3 (10)~TeV and integrated luminosity of 1 (10)~ab$^{-1}$, respectively.

In the fully hadronic $4b$ channel, the dominant backgrounds are Higgs-induced processes with the same or similar visible final states, while non-Higgs contributions are strongly suppressed after cuts. Our baseline selection relies on moderate jet thresholds and, crucially, an angular separation requirement $\Delta R_{jj}>0.4$ to ensure that reconstructed jets are well resolved. Such $\Delta R$ requirement efficiently vetoes collinear and overlapping jet configurations characteristic of soft QCD radiation. It is also essential for suppressing reducible backgrounds such as $b\bar{b}$ + light jet events which come from Higgs or $Z$ decays, where additional (mis-)tagged jets are predominantly generated as soft/collinear shower radiation. 
After applying the preselection rules, the overall $4b$ signal acceptance is at the level of ${\cal O}(10^{-2})$. 

In the $4b$ final state, QCD radiation and jet-combinatorics leave sizeable backgrounds after preselection. We therefore train a BDT-based classifier to help discriminate signal from backgrounds. 
With such ML-based selection, the signal-to-background ratio increases by about a factor of two, and an improvement of $\sim 30\%$ in the expected statistical significance before imposing the resolved $4b$ tagging requirement. After the full selection, the $10~\mathrm{TeV}$ benchmark reaches sensitivity to $\mathrm{BR}(h\to SS\to 4b)$ at the level of $10^{-3}$, while the $3~\mathrm{TeV}$ benchmark with $1~\mathrm{ab}^{-1}$ is limited to substantially weaker, percent-level branching ratios. The reach deteriorates for $m_S$ below about $20~\mathrm{GeV}$, where the resolved-jet requirement $\Delta R_{jj}>0.4$ increasingly removes signal events with collimated $b$ jets. Compared the HL-LHC projection limited to $\mathcal{O}(10^{-1})$ level, both muon collider benchmark scenarios demonstrate clearly high potential for rare exotic Higgs decays with hadronic final states. Finally, since $S$ decays through mixing with the Higgs in the Higgs-portal model, the dominant decay channel of $S$ is $S\to b\bar{b}$ (at $\gtrsim 80\%$ of the times), rendering the limit on BR$(h\to SS)$ only slightly weaker than that of BR$(h\to SS \to 4b)$ in value.

In the $2b2\mu$ final state, the BDT-based ML cut is also applied. In this case, the presence of a narrow dimuon resonance makes the signal straightforward to identify and significantly reduces the combinatorics of jet pairs relative to the fully hadronic mode. 
Accordingly, we include explicit $m_{\mu\mu}$ information only after the ML-based cuts to achieve selection efficiency across benchmarks, so as the $2b$-tagging requirement. In the $10~\mathrm{TeV}$ benchmark, the final sensitivity to $\mathrm{BR}(h\to SS\to 2b2\mu)$ reaches the level of ${\cal O}(10^{-5})$, although in the Higgs-portal model the small $\mathrm{BR}(S\to\mu^+\mu^-)$ dictated by the scalar mixing with the Higgs makes this channel much less competitive than the $4b$ mode in the model-dependent reach of BR$(h\to SS)$.

Overall, the muon collider remains advantageous for low-energy-scale exotic Higgs decays compared with the HL-LHC, especially in hadronic final states for which reducible QCD backgrounds dominate at hadron colliders. For the $h\to SS\to 4b$ topology, the projected reach improves from the HL-LHC level of $\mathcal{O}(10^{-1})$ to about $10^{-2}(10^{-3})$ level at $\sqrt{s}=3(10)~\mathrm{TeV}$, respectively. This gain is not driven by accessing large momentum transfer, since the Higgs production and exotic decays at a muon collider are still electroweak-scale phenomena, but rather by the substantially smaller hadronic background rates. Dedicated Higgs factories are expected to provide an even stronger sensitivity, while a multi-TeV muon collider can be more competitive in measurements that benefit directly from high energy, such as associated $hS$ production when $m_S>m_h/2$.

Looking forward, several extensions could be the next natural steps. Beyond the minimal renormalizable portal, non-renormalizable Higgs operators generically lead to effects that increase with collider energy. Therefore, multi-TeV muon-collider measurements can strengthen the corresponding sensitivity further, as is well known. Within the same $h\to SS$ framework, extending the decay-channel coverage to $\tau$-rich modes such as $2b2\tau$ and possibly $4\tau$ is also well motivated, since these channels can provide useful model-dependent constraints once the considerable $S\to \tau\tau$ branching ratio is taken into account. Such final states are more challenging to analyze than $2b2\mu$ due to multiple neutrinos present and the more demanding tracking and vertexing requirements for $\tau$ reconstruction. Additionally, the investigation of the $h\to SS$ topology can be extended to semi-invisible channels~\cite{Curtin:2013fra}, such as $h\to S + \text{missing energy}$~\cite{Huang:2013ima} driven by dark matter phenomenology~\cite{Draper:2010ew}, and $h\to \text{resonant gauge bosons} + \text{missing energy}$ motivated by neutrino physics~\cite{Graesser:2007yj,deGouvea:2007hks}. Finally, a more refined machine-learning-based analysis framework has the potential to enhance the performance in the low $m_S$ regime and bring the overall constraint power comparable to a Higgs factory. They will be left for future studies.

\section*{Acknowledgements}

JF is supported by the DOE grant DE-SC-0010010. TL is supported by the General Research Fund under Grant No. 16305425, from the Research Grants Council of Hong Kong S.A.R. 

\section*{Appendix}
\label{sec:app}
In this appendix, we provide the yields after the full analysis for all benchmarks with different singlet scalar masses. One could see that for lighter $S$ with $m_S \lesssim 20$ GeV, the signal yields drop significantly due to collimated final-state particles. For larger $m_S$, the yields are similar for different masses in a given channel with fixed muon collider setup. 

\begin{table}[htbp]
\centering
\resizebox{\textwidth}{!}{%
\begin{tabular}{lcccccc}
\toprule
$m_S$ & 15 GeV  & 20 GeV & 30 GeV & 40 GeV & 50 GeV & 60 GeV \\
\midrule
\multicolumn{7}{l}{\textbf{Signal}} \\
$h \to SS \to 4b$  & $\leq 2.5\times 10^{2}$ BR & $2.4 \times 10^3$ BR & $1.1 \times 10^4$ BR & $1.3 \times 10^4$ BR & $8.1 \times 10^3$ BR & $1.3 \times 10^4$ BR \\
\midrule
\multicolumn{7}{l}{\textbf{Background}} \\
$h \to ZZ^* \to 4b$ & 0.19 & 0.29 & 4.3 & 7.0 & 7.0 &16.5\\
$h \to 4b$ & 0.053 & 0.053 & 0.87 & 2.9 & 2.3 & 1.7\\
$2b$ &$\leq 0.76$ & $\leq 0.76$ & $\leq 0.76$ & $\leq 0.76$ & $\leq 0.76$ & $\leq 0.76$ \\
$4b$ & $\leq 0.44$ & $\leq 0.44$ & 1.0  & 1.5& 1.5 & 3.3 \\
\bottomrule
\end{tabular}
}
\caption{Final yields of signals and backgrounds for different $m_S$ benchmarks in the $4b$ channel, at a 10 TeV muon collider with 10 ab$^{-1}$ data.}
\label{tab:4byield10TeV}
\end{table}

\begin{table}[htbp]
\centering
\resizebox{\textwidth}{!}{%
\begin{tabular}{lcccccc}
\toprule
$m_S$ & 15 GeV  & 20 GeV & 30 GeV & 40 GeV & 50 GeV & 60 GeV \\
\midrule
\multicolumn{7}{l}{\textbf{Signal}} \\
$h \to SS \to 4b$  & $\leq 3.0\times 10^{1}$  BR & $2.7 \times 10^2$ BR & $1.1 \times 10^3$ BR & $7.0\times 10^2$ BR & $8.5\times 10^2$ BR & $7.5\times 10^2$ BR \\
\midrule
\multicolumn{7}{l}{\textbf{Background}} \\
$h \to ZZ^* \to 4b$ & $\leq 0.01$  & $\leq 0.01$  & 0.11 & 0.16 & 0.15 &0.42\\
$h \to 4b$ & $\leq 0.01$  & $\leq 0.01$ & 0.06  & 0.15 & 0.13 & $0.06$\\
$2b$ & $\leq 0.01$ & $\leq 0.05$ & 0.07 & 0.11 & $\leq 0.07$ & $\leq 0.07$ \\
$4b$ & $\leq 0.01$ & $\leq 0.01$ & 0.03  & 0.09& $0.08$ & $0.05$ \\
\bottomrule
\end{tabular}
}
\caption{Final yields of signals and backgrounds for different $m_S$ benchmarks in the $4b$ channel, at a 3 TeV muon collider with 1 ab$^{-1}$ data.}
\label{tab:4byield3TeV}
\end{table}

\begin{table}[htbp]
\centering
\resizebox{\textwidth}{!}{%
\begin{tabular}{lcccccc}
\toprule
$m_S$ & 15 GeV  & 20 GeV & 30 GeV & 40 GeV & 50 GeV & 60 GeV \\
\midrule
\multicolumn{7}{l}{\textbf{Signal}} \\
$h \to SS \to 4b$  & $1.7\times 10^{4}$  BR & $9.0 \times 10^4$ BR & $2.2 \times 10^5$ BR & $3.1\times 10^5$ BR & $2.2\times 10^5$ BR & $2.1\times 10^5$ BR \\
\midrule
\multicolumn{7}{l}{\textbf{Background}} \\
$h \to 2b2\mu$ & 1.0  &1.1 & 1.5 & 0.96 & 1.2 &1.9\\
$2b2\mu$ & 0.74  & 1.0 & 0.83  & 1.2 & 0.83 & 1.6\\
\bottomrule
\end{tabular}
}
\caption{Final yields of signals and backgrounds for different $m_S$ benchmarks in the $2b2\mu$ channel, at a 10 TeV muon collider with 10 ab$^{-1}$ data.}
\label{tab:2b2myield10TeV}
\end{table}

\begin{table}[htbp]
\centering
\resizebox{\textwidth}{!}{%
\begin{tabular}{lcccccc}
\toprule
$m_S$ & 15 GeV  & 20 GeV & 30 GeV & 40 GeV & 50 GeV & 60 GeV \\
\midrule
\multicolumn{7}{l}{\textbf{Signal}} \\
$h \to SS \to 4b$  & $1.6\times 10^{3}$  BR & $8.0 \times 10^3$ BR & $2.4 \times 10^4$ BR & $2.8\times 10^4$ BR & $2.5\times 10^4$ BR & $2.3\times 10^4$ BR \\
\midrule
\multicolumn{7}{l}{\textbf{Background}} \\
$h \to 2b2\mu$ & 0.041  &0.075 & 0.074 & 0.097 & 0.12 &0.074\\
$2b2\mu$ & 0.10  & 0.10 & 0.15  & 0.27 & 0.28 & 0.21\\
\bottomrule
\end{tabular}
}
\caption{Final yields of signals and backgrounds for different $m_S$ benchmarks in the $2b2\mu$ channel, at a 3 TeV muon collider with 1 ab$^{-1}$ data.}
\label{tab:2b2myield3TeV}
\end{table}

\bibliography{Ref}
\end{document}